\documentclass[journal,10pt,]{IEEEtran}
\normalsize
\usepackage[T1]{fontenc}% optional T1 font encoding
\usepackage[pdftex]{graphicx}     

\usepackage{cite}
\ifCLASSINFOpdf
\else
\fi
\usepackage{amsmath}
\usepackage[cmintegrals]{newtxmath}

\usepackage[caption=false,font=footnotesize,labelfont=sf,textfont=sf]{subfig}
\usepackage{stfloats}

\usepackage{pgfplots}
\pgfplotsset{compat=1.13}

\usepackage{tikz}

\begin{document}
\title{Physical Layer Identification based on Spatial-temporal Beam Features for Millimeter Wave Wireless Networks}

\author{Sarankumar~Balakrishnan,
		Shreya Gupta,
        Arupjyoti~Bhuyan,
        Pu Wang,
        Dimitrios Koutsonikolas,
        and~Zhi~Sun% <-this % stops a space
\thanks{S.~Balakrishnan, S.~Gupta and Z.~Sun are with the Department of Electrical Engineering, University at Buffalo, Buffalo, NY 14260 USA (e-mail: sarankum@buffalo.edu; shreyagu@buffalo.edu; zhisun@buffalo.edu).}% <-this % stops a space
\thanks{A.~Bhuyan is with Idaho National Laboratory (INL), Idaho Falls, ID, 83402 USA (e-mail: arupjyoti.bhuyan@inl.gov).}% <-this % stops a space
%\thanks{}% <-this % stops a space
\thanks{P.~Wang is with the Dept. of Computer Science, University of North Carolina at Charlotte, Charlotte, NC, 28223 USA (e-mail: pu.wang@uncc.edu).}% <-this % stops a space
\thanks{D.~Koutsonikolas is with the Deptartment of Computer Science and Engineering, University at Buffalo, Buffalo, NY 14260 USA (e-mail: dimitrio@buffalo.edu).}% <-this % stops a space
\thanks{$^\dagger$Work supported through the INL Laboratory Directed Research \& Development (LDRD) Program under DOE Idaho Operations Office Contract DE-AC07-05ID14517.}
}

\maketitle

\begin{abstract}
With millimeter wave (mmWave) wireless communication envisioned to be the key enabler of next generation high data rate wireless networks, security is of paramount importance. While conventional security measures in wireless networks operate at a higher layer of the protocol stack, physical layer security utilizes unique device dependent hardware features to identify and authenticate legitimate devices. In this work, we identify that the manufacturing tolerances in the antenna arrays used in mmWave devices contribute to a beam pattern that is unique to each device, and to that end we propose a novel device fingerprinting scheme based on the unique beam pattern of different codebooks used by the mmWave devices. Specifically, we propose a fingerprinting scheme with multiple access points (APs) to take advantage of the rich spatial-temporal information of the beam pattern. We perform comprehensive experiments with commercial off-the-shelf mmWave devices to validate the reliability performance of our proposed method under various scenarios. We also compare our beam pattern feature with a conventional physical layer feature namely power spectral density feature (PSD). To that end, we implement PSD feature based fingerprinting for mmWave devices. We show that the proposed multiple APs scheme is able to achieve over $99\%$ identification accuracy for stationary LOS and NLOS scenarios and significantly outperform the PSD feature fingerprinting method. For mobility scenarios, the overall identification accuracy is $96\%$. In addition, we perform security analysis of our proposed beam pattern fingerprinting system and PSD fingerprinting system by studying the feasibility of performing impersonation attacks. We design and implement an impersonation attack mechanism for mmWave wireless networks using state-of-the-art 60 GHz software defined radios. We discuss our findings and their implications on the security of the mmWave wireless networks.
\end{abstract}

\begin{IEEEkeywords}
Millimeter wave, physical layer security, RF fingerprinting, 802.11ad, 5G.
\end{IEEEkeywords}

% \fi
\IEEEpeerreviewmaketitle

\section{Introduction}
%\IEEEPARstart{T}{his}
Increasing demands for higher data rates and the availability of wide bandwidth at higher frequency spectrum makes mmWave communication attractive for next generation wireless systems. MmWave communication is seen as an enabling technology to multi-Gigabit WLANs, wireless display, cable free connection, virtual reality, to name a few. The current 60 GHz WLAN IEEE standard 802.11ad and the upcoming standards like IEEE 802.11ay and 5G NR for cellular networks use mmWave for communication.   
\par With the proliferation of mmWave wireless communication, enormous amount of data will be transmitted over wireless. It is estimated that by 2020 there will be 38 billion connected devices with more than $60\%$ of global mobile traffic through WiFi \cite{WiFialliance}. The majority of this traffic will be driven by next generation mmWave wireless networks such as 802.11 ad/ay. Hence security is critical for mmWave wireless networks. Existing security protocols for wireless standards including mmWave are implemented at the software level and are traditionally cryptographic based schemes such as WPA, WPA2-PSK and they are proven to be vulnerable to several attacks such as DoS attack \cite{raymond2008denial}, injection attack \cite{ohigashi2009practical}, spoofing attack \cite{kannhavong2007survey} and man-in-the-middle (MITM) attack. 
Also, mmWave communication is inherently considered to be secure due to the directionality of the antenna beams and attacks such as MITM is considered to be difficult. Authors in \cite{Steinmetzer2018BeamStealingIT} have demonstrated the potential vulnerability in the beam searching mechanism currently adopted in the IEEE 802.11ad standard. Through firmware modifications of the TALON AD7200 router, they successfully sent a forged sector sweep feedback frame to the legitimate destination device to divert the beam from that device to the attacker. \par
Recently several Intel WLAN cards including the 60 GHz Intel Tri-Band Wireless AC-18265 card that we use in our experiments were exposed to frame replay vulnerability \cite{IntelFrameReplay}. In this attack, an attacker that successfully establishes channel based MITM can potentially replay frames to the destination. This vulnerability allowed the replayed frames to pass on to the operating system as new frames, potentially compromising the integrity of the already transmitted legitimate frames.
\par Attacks such as \cite{Steinmetzer2018BeamStealingIT} and \cite{IntelFrameReplay} expose the security vulnerability of present mmWave wireless networks and could have severe impact on the security of the system as they can not be prevented using higher layer security protocols \cite{Steinmetzer2018BeamStealingIT},\cite{IntelFrameReplay}. Hence there exists a strong need for security measures in addition to the conventional security methods that are currently used. Such attacks could be mitigated by verifying the authenticity of the frames originating from the imposter devices at the physical layer by using physical layer identification techniques.  
\par Recently physical layer security has become a promising solution to address the aforementioned security issues and augment the security of wireless systems. Wireless waveforms transmitted by the device are stamped with unique features that originate in the physical layer of the transmitter that could be potentially used to identify and authenticate devices. Such unique features are generated by the imperfections along the hardware chain of the transmitter. The features introduced by the hardware are difficult to forge unlike software based security schemes and could be used in conjunction with the existing higher layer security mechanisms if they are reliable and stable \cite{wang2016wireless}. 
\par The RF fingerprints can origin anywhere along the transmitter chain like clock jitter, I/Q offset due to imbalance between I and Q branch, oscillators, synthesizer for up conversion, DAC sampling process, non-linearity of the amplifiers, phase noise of the phase shifters and fabrication process tolerance inherent to antenna elements. Current RF physical layer features are either 1) data dependent like transient and preamble based features that are prone to signal replay attack, or 2) low-dimensional like modulation based features which can be easily forged and have limited capacity in terms of number of devices that can be enrolled, which makes them weak and limited in practical usage. 
\par Existing works on RF fingerprinting were proposed for conventional sub-6 GHz wireless band and physical layer security schemes for mmWave communications remain largely unexplored. The propagation characteristics of mmWave signals add a unique dimension to the fingerprinting problem. To overcome propagation losses at higher frequencies, mmWave devices uses beamforming enabled by antenna arrays. Now the natural question arises: \textit{Can the antenna arrays in mmWave devices generate unique fingerprints that could be used to reliably identify and authenticate mmWave devices?} The antenna arrays, owing to the errors and tolerances in manufacturing processes (Sec \ref{sec:feature origin}), generate distinct beam patterns that are unique among devices. Typically mmWave devices use a set of beam patterns and find the best one to use through a process known as beam searching. The mobility of the user or device orientation change exposes different angular views of these beam patterns. We propose to leverage this spatial-temporal charactersitic of the beam pattern used by the user device. We show that these device dependent beam patterns can be learned and reliably employed for device identification. To the best of our knowledge, no work exists in the literature that utilizes the spatial beam patterns of the mmWave antenna arrays for RF fingerprinting. The proposed beam pattern feature has the following properties: 1) data independent as beam pattern does not depend on the signal being transmitted, 2) high dimensional and 3) resilient to impersonation attacks.   
\subsection{Contribution}
Our contributions can be summarized as follows:
\begin{enumerate}
\item We identify that the fabrication process of the antenna array and the phase shifters used by mmWave devices introduce unique variations of beam patterns among the devices. Motivated by this observation, we propose a novel spatial-temporal beam feature for mmWave fingerprinting based on the beam patterns swept by the mmWave device during the beam searching process. To exploit the rich spatial-temporal feature in the beam pattern sweep during the beam searching phase, we propose a multiple APs architecture for RF fingerprinting mmWave devices and provide an optimal deployment strategy for the APs. We analyze the potential user capacity of the proposed beam pattern feature through COMSOL \cite{comsol} simulations and find that the beam pattern feature can support large number of users in the order of thousands.  
\item We demonstrate the reliability and robustness of the proposed spatial-temporal beam feature through extensive experiments with commercial mmWave devices. In addition, we compare our proposed feature with an existing conventional RF feature. To that end, we implement PSD based device fingerprinting scheme for mmWave devices. Our proposed fingerprinting system achieves a very high identification accuracy under stationary LOS and NLOS scenarios when compared to the conventional RF feature. We also studied the impact of mobility on the performance of beam pattern feature. Moreover, our proposed fingerprinting scheme does not need additional signal processing or hardware as opposed to the conventional feature which typically requires complex signal processing and expensive hardware owing to the high bandwidth of the mmWave signal.
\item We perform security analysis of our proposed mmWave feature by studying its resilience to impersonation attacks. To that end, we implement impersonation attack on the fingerprinting system using state-of-the art mmWave software defined radio (SDR) and show that with multiple APs, impersonation attacks on the beam pattern feature can be successfully thwarted. On the other hand, we show that the conventional feature is vulnerable to impersonation attacks thus severely limiting its practical usage.
\end{enumerate}
\subsection{Related Work} \label{sec:related work}
Several works exist in the literature that investigate the RF fingerprints for physical layer identification and classification. Authors in \cite{danev2009transient} use the turn-on and turn-off transient part of the RF signal for fingerprinting. However, acquiring the transient part of the signal requires expensive high end signal acquisition set up which makes it difficult for many applications that require low cost fingerprinting techniques. The approach in \cite{brik2008wireless} uses frequency error, SYNC correlation, I/Q offset, magnitude error and phase error features extracted from IEEE 802.11 frames and is shown to achieve classification accuracy of over $99\%$. Even though such features are comparatively easier to extract, they are proven to be vulnerable to signal and feature replay attacks \cite{danev2010attacks}. The authors in \cite{danev2009physical} study the physical layer identification of RFID devices using modulation based features and spectral features. They show that their proposed RFID features achieves a $0\%$ classification error rate. 
\par Very limited work exists in the literature that studies the attacks on physical layer fingerprinting methods. The work in \cite{danev2010attacks} experimentally demonstrates signal replay and feature replay attacks on IEEE 802.11 devices. It shows that low dimensional features such as frame frequency offset or IQ origin offset can be replayed using high end arbitrary waveform generators.
\par All these works on RF fingerprinting are proposed for sub-6 GHz wireless technologies and to the best of the author knowledge no work exists in the literature for fingerprinting commercial mmWave devices. Also the above mentioned transient and modulation based fingerprints are constrained in feature space, i.e., they rely on low dimensional feature space leading to constraints on the number of devices successfully identified by those features. In this work, distinct from the above mentioned RF fingerprinting schemes where the spatial features are typically not considered, we propose a novel high dimensional RF fingerprinting scheme based on the spatial signatures of the beam pattern used by the mmWave devices. 
\subsection{Organization}
We discuss the proposed mmWave fingerprinting system architecture and feature origin in Sec. \ref{sec:system design}. In Sec. \ref{sec:system implementation}, we describe our system implementation and testbed set up. In Sec. \ref{sec:experiments}, we present comprehensive experimental results for various practical scenarios and discuss the findings. Section \ref{sec:attacks} discusses impersonation attacks on the proposed mmWave feature. Section \ref{sec:conclusions} discusses conclusions. 

\section{System Design} \label{sec:system design}
This section describes the overall system architecture, source of proposed beam pattern feature, AP deployment strategy and beam pattern feature extraction protocol.
\subsection{System Architecture and Operational Framework}
\begin{figure}[!ht]
  \centering
  \includegraphics[trim={2cm 1cm 8cm 6cm},clip,width=7cm]{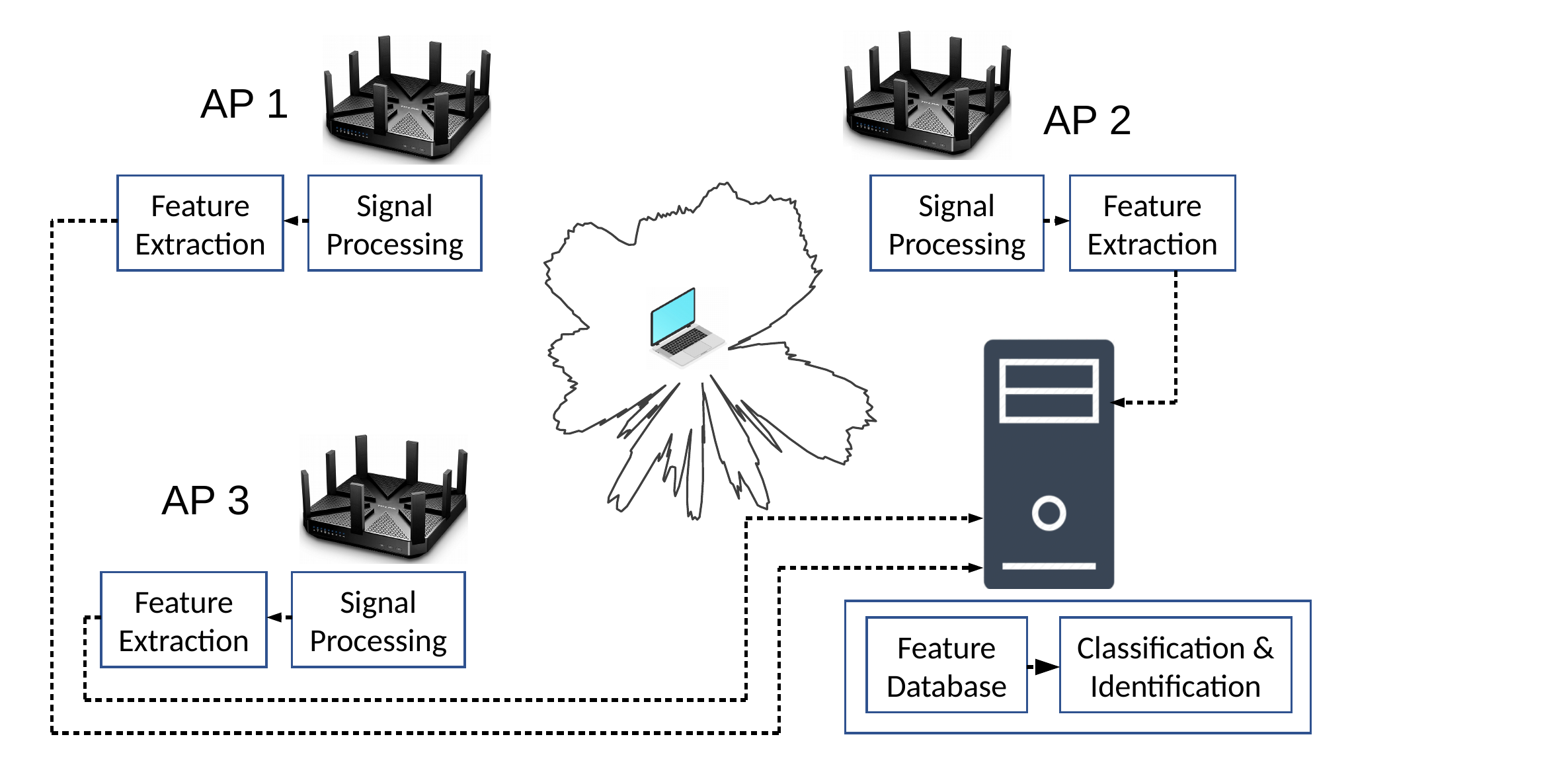}
  \caption{MmWave fingerprinting system architecture. \label{fig:system architecture}}
\end{figure}
We consider a mmWave wireless network with multiple APs and clients deployed in an indoor environment as shown in Fig. \ref{fig:system architecture}. The APs and clients follow the mmWave standard (e.g., 802.11ad) and perform beam searching to establish directional communication with each other. The APs utilize a unique hardware dependent beam pattern feature (Sec. \ref{sec:feature origin}) to authenticate the clients joining the network. The beam pattern feature is extracted during the beam searching phase that already exists in the mmWave standards such as 802.11ad \cite{80211adstandard}, 802.11ay \cite{da2018beamforming} and 5G NR \cite{lien20175g} and does not introduce any additional protocol and signal processing overhead. The APs are placed in an optimal position in the deployment area using the method described in Sec. \ref{sec:AP deployment}. All the APs are connected to a back-end server through a backhaul network or Gigabit Ethernet. The back-end server runs the classification and identification algorithm to authenticate the clients. The system operates in a two-stage process: \textit{1) learning stage} and \textit{2) identification stage}. During the learning stage, devices are enrolled and beam pattern feature databases for the enrolled devices are built. Each device is given an identity label $L_i,i=1,2,\cdots,N$ with $N$ being the number of clients. The APs initiate the beam searching mechanism (sector level sweep in 802.11ad) periodically and also triggered by device mobility, extract the beam pattern feature vector and communicate the feature vector and device identity label $L_{i}$ to the backend server. The learning stage is a one-time process and is completed before the client devices are authenticated and admitted to the system. The backend server builds the beam pattern feature database for each enrolled client $L_i$ and performs learning/training using the learning method discussed in Sec. \ref{sec:cnn}. During the identification stage, the device that intends to join the network, performs beam searching with the APs. Each AP extracts the beam pattern feature vector of the device to be identified and the backend server verifies it against the feature of the claimed identity of the device. 

\subsection{Phased Array Beam Pattern Feature: A Closer Look} \label{sec:feature origin}
In this section we take a closer look at the antenna arrays found in commercial mmWave devices to identify the source of fingerprint due to beam pattern variations. Without loss of generality, assume a 2D planar antenna array which is popularly used for mmWave applications. The beam pattern of the antenna array is given by
\begin{equation}
\label{eq:beam pattern}
f(\theta,\phi) = |f_{el}(\theta,\phi)\sum_{m=0}^{M_{x}-1}\sum_{n=0}^{M_{y}-1}w_{m,n,k}e^{j2\pi(md_xacos\phi+nd_yasin\phi)}|^2,
\end{equation}
where $a=\frac{sin\theta}{\lambda}$ and $f_{el}(\theta,\phi)$ is the radiation field of individual antenna element. $\lambda$ is the wavelength, $M_x$ and $M_y$ are the number of antenna elements along x-axis and y-axis respectively. $w_{m,n,k}=\alpha_{m,n} e^{j\delta_{m,n}}$ is the complex excitation of the $m,n$th element to form the $k$th beam pattern with $\alpha_{mn}$ and $\delta_{mn}$ being the amplitude and phase excitations of $m,n$th element.
From (\ref{eq:beam pattern}), the beam pattern of the antenna array is a function of the radiation field of individual elements, amplitude and phase applied to each of the elements and the array geometry. In addition to these, the beam pattern of the array is also affected by the manufacturing tolerances and manufacturing errors that arise due to masking, etching and dielectric constant tolerances during antenna array fabrication. 
\par The first source of error in antenna array fabrication is due to the dielectric properties of the substrate used in the antenna. 
The antenna patch resonance frequency depends on the relative permittivity of the material used for substrate and is given by $f_c \sim \frac{c}{2*L*\sqrt(\epsilon_r)}$ where $c$ is speed of light in vacuum, $L$ is the patch length and $\epsilon_r$ is the relative permittivity. Some of the popular substrates used for mmWave antenna are Rogers\textsuperscript{\textregistered}RO3003, RO3203 \cite{Rogers:3000}, duroid 5880 \cite{duroid5880}, and LTCC Ferro A6-S \cite{FerroA6-S}. The relative permittivity $\epsilon_r$ of these subtrates has a tolerance value that deviates from the specified $\epsilon_r$. E.g., Ferro A6-S substrate has a $\epsilon_r$ of $5.9\pm0.2$ \cite{FerroA6-S}. 
\par A second source of variations in the beam patterns is the dimensional tolerance of the antenna fabrication process. The manufacturing error increases considerably at higher frequencies due to reduced antenna size and also due to increasing substrate dielectric $\epsilon_r$ (antenna patch width decreases with increase in $\epsilon_r$). Due to the shrinkage in dimension during the fabrication process and the errors that depend on the tolerance of the process used to fabricate, the final dimensions of the antenna are not the same across different batches. The Ferro A6-S substrate typically used for mmWave antenna fabrication has a x-y shrinkage as high as $15\%\pm2$ \cite{FerroA6-S}.
\par As shown in (\ref{eq:beam pattern}), phase shifts are applied to each element of the antenna array to steer the beam toward intended direction. The phase shifters used for beamforming in commercial mmWave devices are typically low cost, low resolution with high phase errors. The $i$th phase shifter has a phase deviation error of $\pm \Delta \delta_i$ from the actual phase value and the error is independent among the phase shifters used in different antenna elements. If $\theta_i$ is the ideal phase shift, the actual phase shift applied to the $i$th element is $\theta_i \pm \Delta \delta_i$. For 15 to 26 GHz frequency range, the phase shifter in \cite{koh20070} introduces an rms phase error of $6.5^\circ-13^\circ$ and for 57-64GHz range, the 5-bit phase shifter in \cite{li201360} has a rms phase error of $<10^{\circ}$. The 3-bit phase shifter in \cite{elkind201857} for 60 GHz has a maximum phase error of $10.4^{\circ}$ and rms error of $5.7^{\circ}$. 
\par All these errors due to tolerances associated with antenna fabrication process as well as the phase shifters, introduce unique hardware dependent beam pattern variation among antenna arrays even with the same architecture and geometry. 
\paragraph*{\textbf{Antenna Array Design and Simulation}}
\label{sec:array design}
\begin{figure}[ht] 
\begin{minipage}[b]{0.43\linewidth}
\centering
\includegraphics[trim={0cm 0cm 0cm 0cm},clip,width=0.5\textwidth]{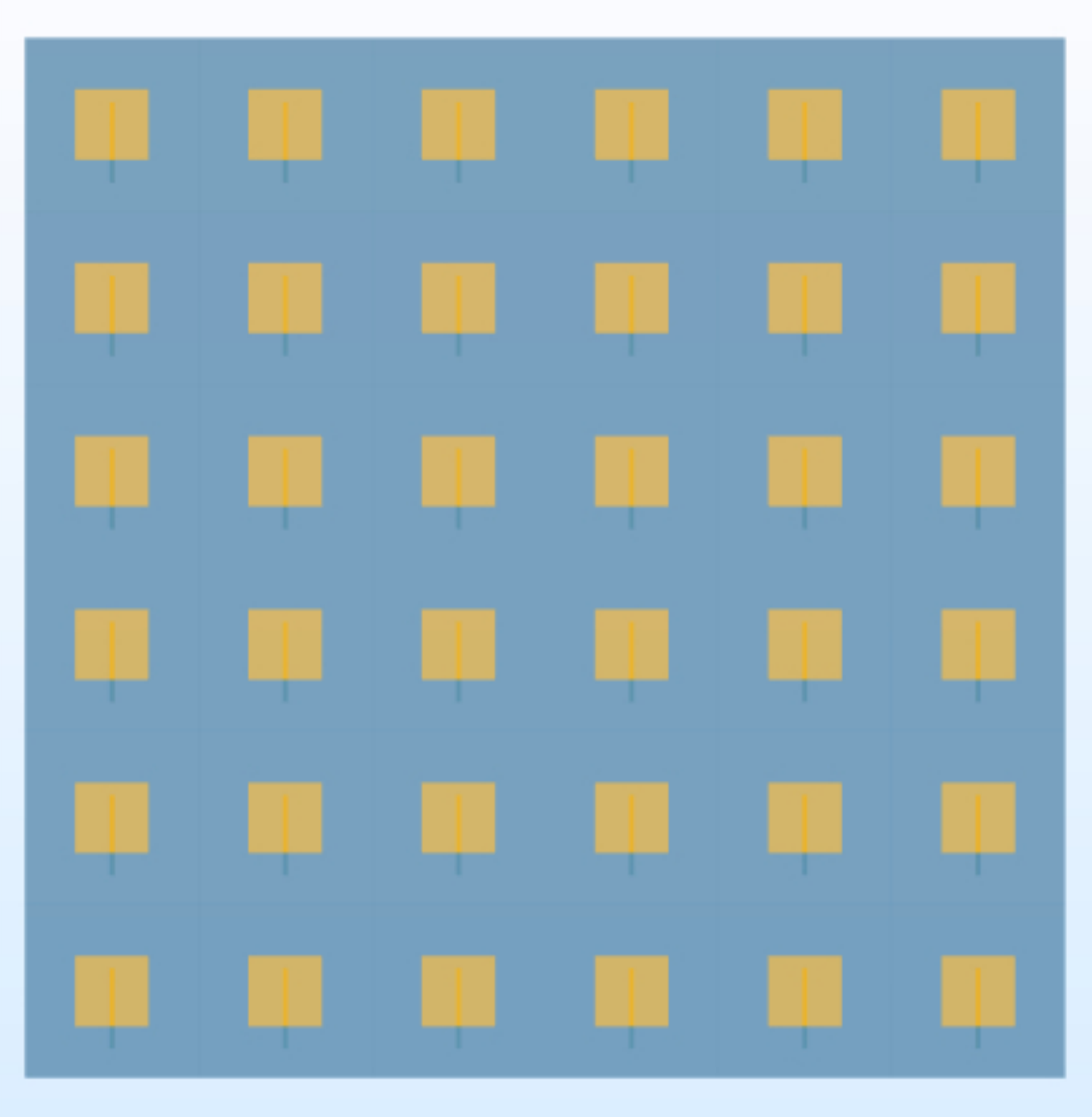}
\caption{$6\times 6$ 30 GHz antenna array designed using COMSOL.}
\label{fig:comsol_design}
\end{minipage}
  \hspace{1em}
 \begin{minipage}[b]{0.43\linewidth}
 \centering
  \includegraphics[trim={0cm 0cm 0 0cm},clip,width=1\textwidth]{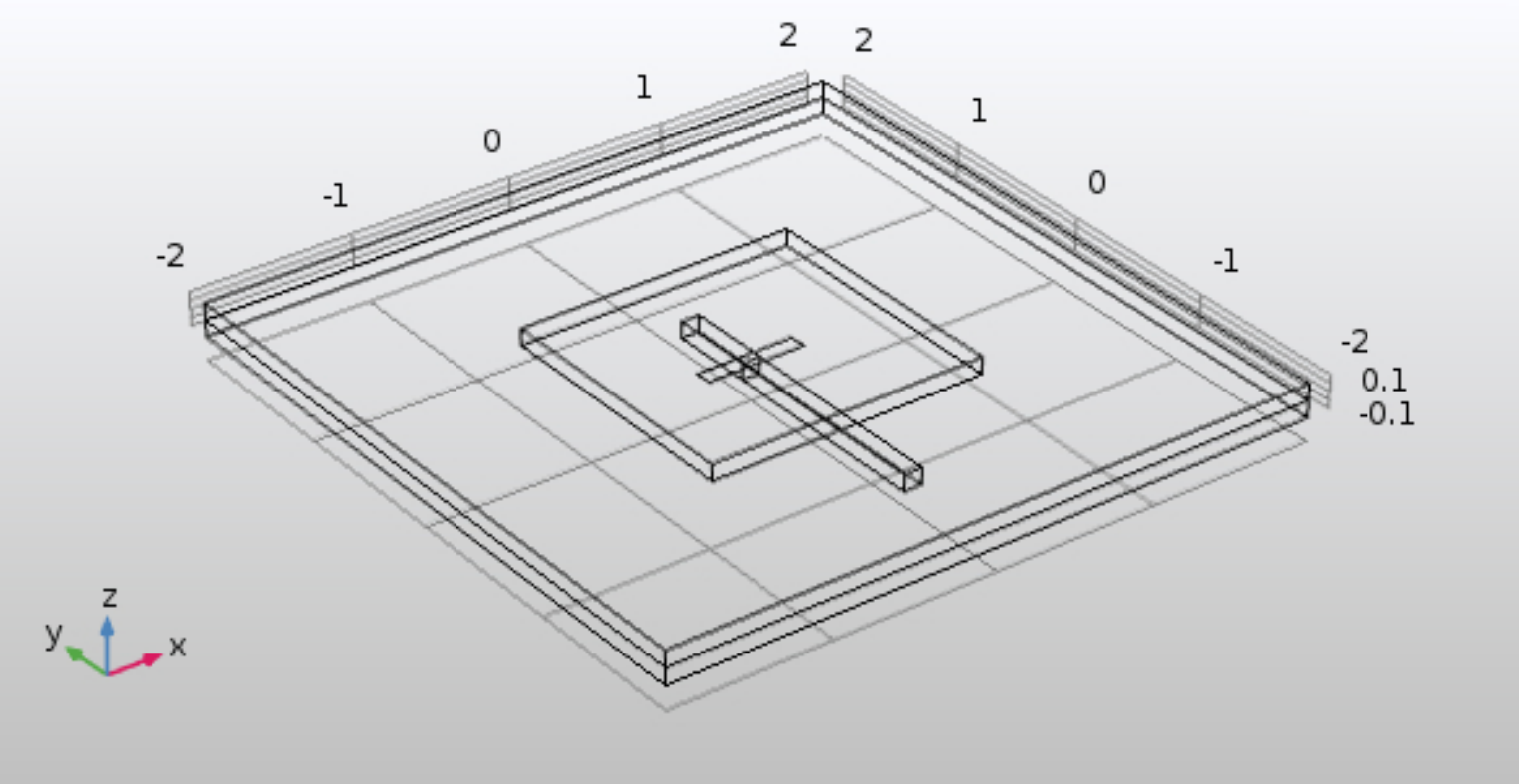}
  \caption{Single element of $6\times 6$ antenna array.}
  \label{fig:comsol_singlepatch}
  \end{minipage}
\end{figure}

\begin{figure*}[htbp!]
%\centering
\subfloat[\label{Fig:beam pattern source dielectric}
]{\includegraphics[trim={10.9cm 8cm 6cm 8cm},clip,width=0.1\textwidth,angle=90]{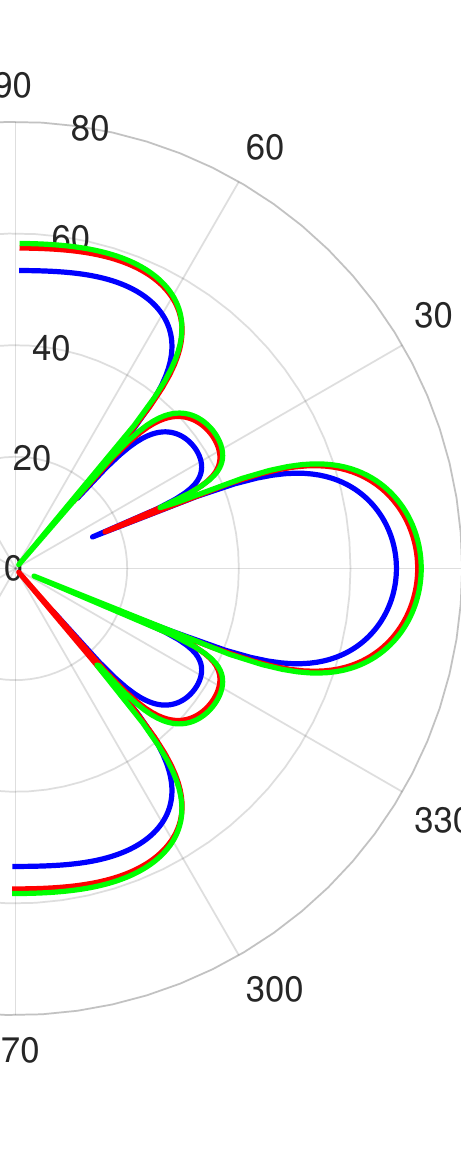}}
\subfloat[\label{Fig:beam pattern source dimension}]{\includegraphics[trim={11.2cm 5cm 3.3cm 6cm},clip,width=0.1\textwidth,angle=90]{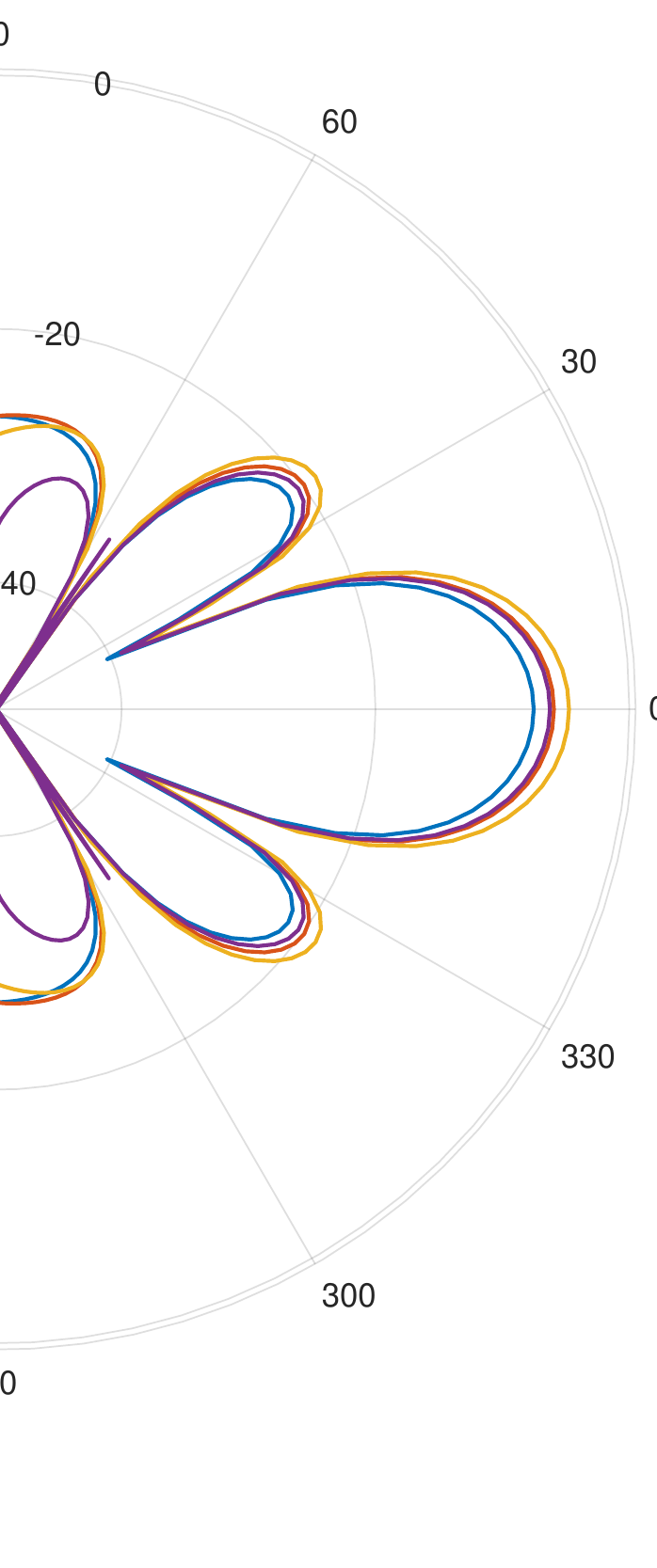}}
\subfloat[\label{Fig:beam pattern element spacing}]{\includegraphics[trim={11.2cm 5cm 3.3cm 6cm},clip,width=0.1\textwidth,angle=90]{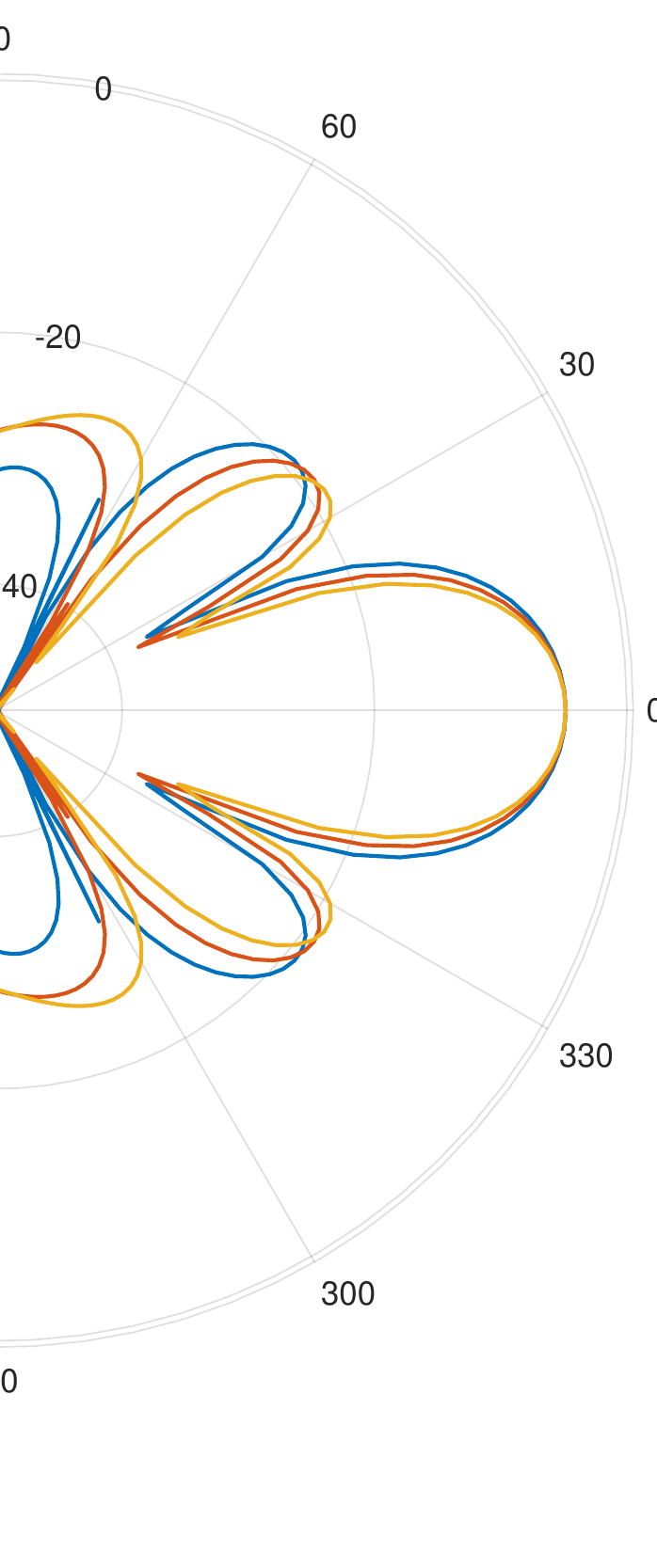}}
\subfloat[\label{Fig:beam pattern source phaseshifter}
]{\includegraphics[trim={11.2cm 5cm 3.3cm 6cm},clip,width=0.1\textwidth,angle=90]{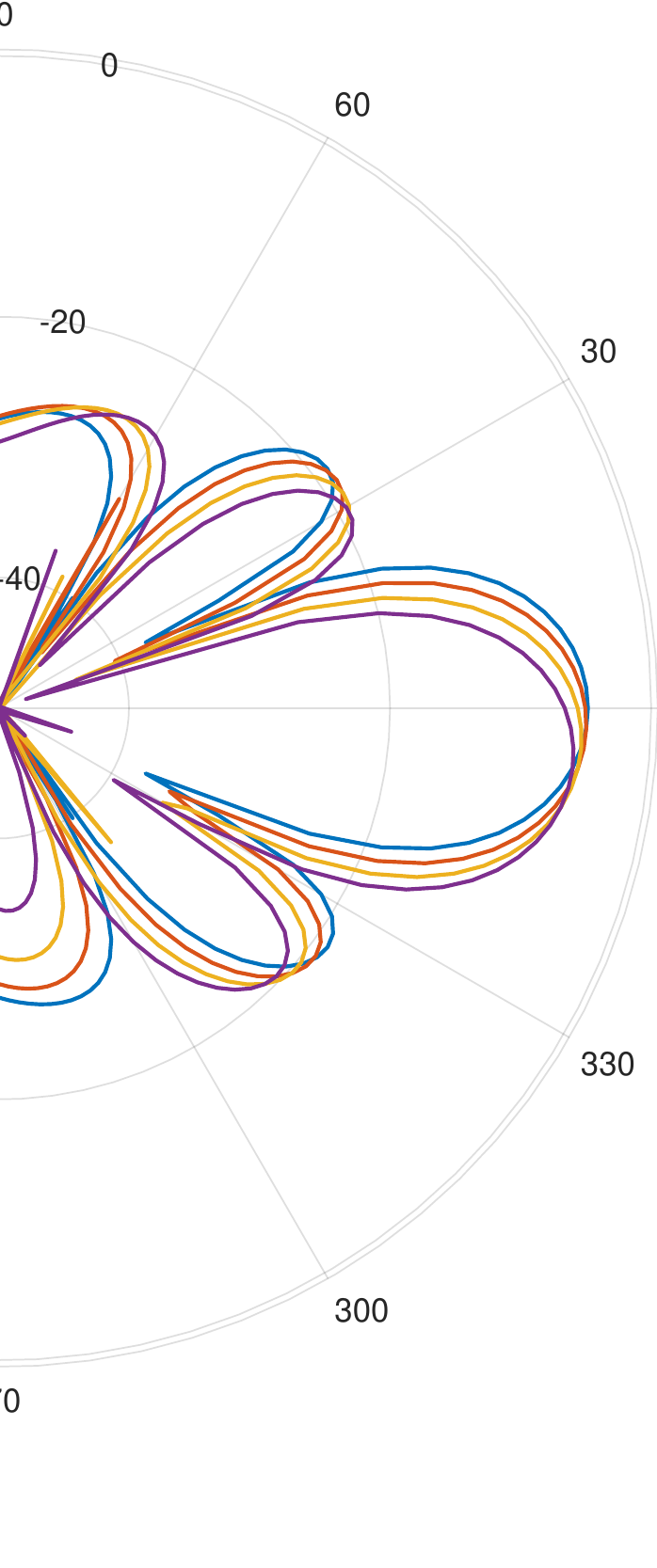}}
\caption{Beam pattern variation due to the tolerance of (a) substrate dielectric, (b) antenna dimension, (c) inter element spacing and (d) phase shifters.}
\label{fig:beam pattern source}
\end{figure*}
To further understand the effect of the antenna manufacturing tolerances and phase shifter errors on the beam pattern variations of mmWave antenna array, we design a $6 \times 6$ slot-coupled microstrip patch antenna array using COMSOL as shown in Fig. \ref{fig:comsol_design} and Fig. \ref{fig:comsol_singlepatch}. The resonant frequency of the array is $30$ $GHz$. The dimensions of the individual antenna elements are as follows: The substrate dimensions are $4mm \times 4mm$. The patch length and width are $1.7mm$ and $1.63mm$, respectively. The slot width and length are $0.1mm$ and $0.605mm$, respectively. The feedline is of width $0.11mm$ and the length of the extended feedline is $0.52mm$. The thickness of the patch and feed substrate is $0.1mm$. The dielectric constant of the substrate is $5.9$. The spacing between the antenna element is $4mm$ along the X and Y directions. The phase applied to the individual antenna elements is $0^{\circ}$. As discussed previously, the variations of the antenna beam pattern feature come from the manufacturing process of the antenna array. Hence, in our simulation, we vary the above mentioned antenna array properties within tolerance limits \cite{natarajan2011fully}, \cite{seki200560}, \cite{FerroA6-S} to understand the effect of antenna fabrication process variations on the beam pattern. We consider a Ferro A6-S substrate with $\epsilon_r = 5.9 \pm 0.2$. The X, Y shrinkage of the LTCC process is assummed to be $15\%$ and the phase shifter error is $5^{\circ}$. Fig. \ref{fig:beam pattern source} shows the beam pattern variations due to the substrate dielectric tolerance, dimension shrinkage, antenna element spacing variation and phase shifter errors. We can see that the tolerances associated with the materials and fabrication process introduce variations in the beam pattern of the antenna array. 

\paragraph*{\textbf{Feature uniqueness}}
For practical usage of device fingerprinting, the proposed beam pattern feature must be unique among devices and scalable. From our simulated beam pattern in Fig. \ref{fig:beam pattern source}, we see that errors due to tolerances of the antenna manufacturing process and phase shifters result in variation of the beam pattern among different antenna arrays. The variations are found to be as high as $3.2dB$. Through our experiments in Sec. \ref{sec:experiments}, we show through classification accuracy that these unique beam pattern variations among devices of same manufacturer and across manufacturers can be learned with high accuracy. Furthermore, we also provide analysis on the maximum number of users that can be supported by our proposed beam pattern feature.
\paragraph*{\textbf{Feature stability}}
\begin{figure}[ht]
\begin{minipage}[b]{0.43\linewidth}
\centering
\includegraphics[trim={11.2cm 5.8cm 2cm 6cm},clip,angle=90,width=1\textwidth]{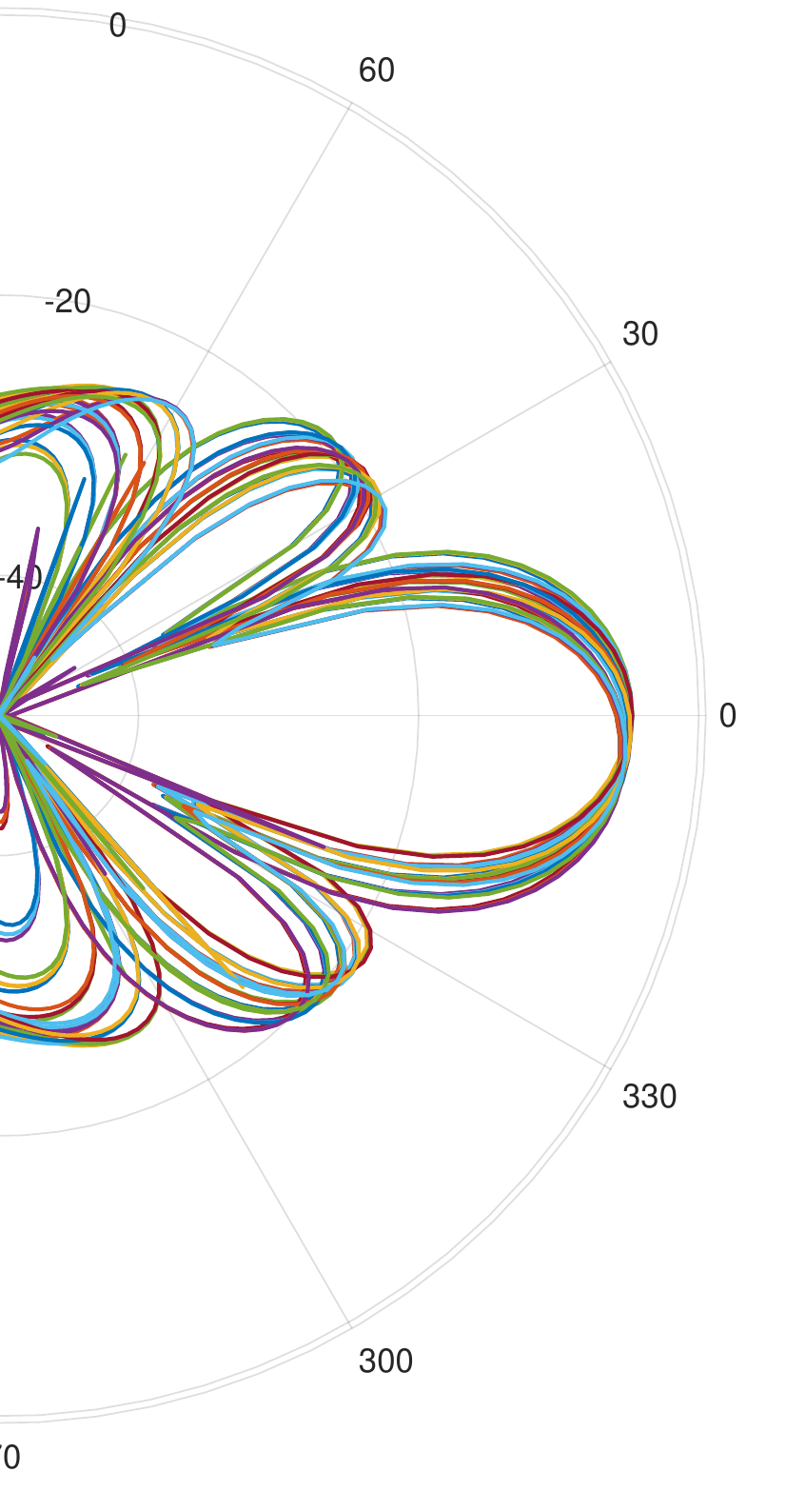}
\caption{Beam pattern variation due to antenna fabrication and phase shifter errors.}
\label{fig:comsol_beampattern}
\end{minipage}
  \hspace{1em}
 \begin{minipage}[b]{0.43\linewidth}
 \centering
  \includegraphics[trim={1cm 6cm 2cm 7cm},clip,width=1\textwidth]{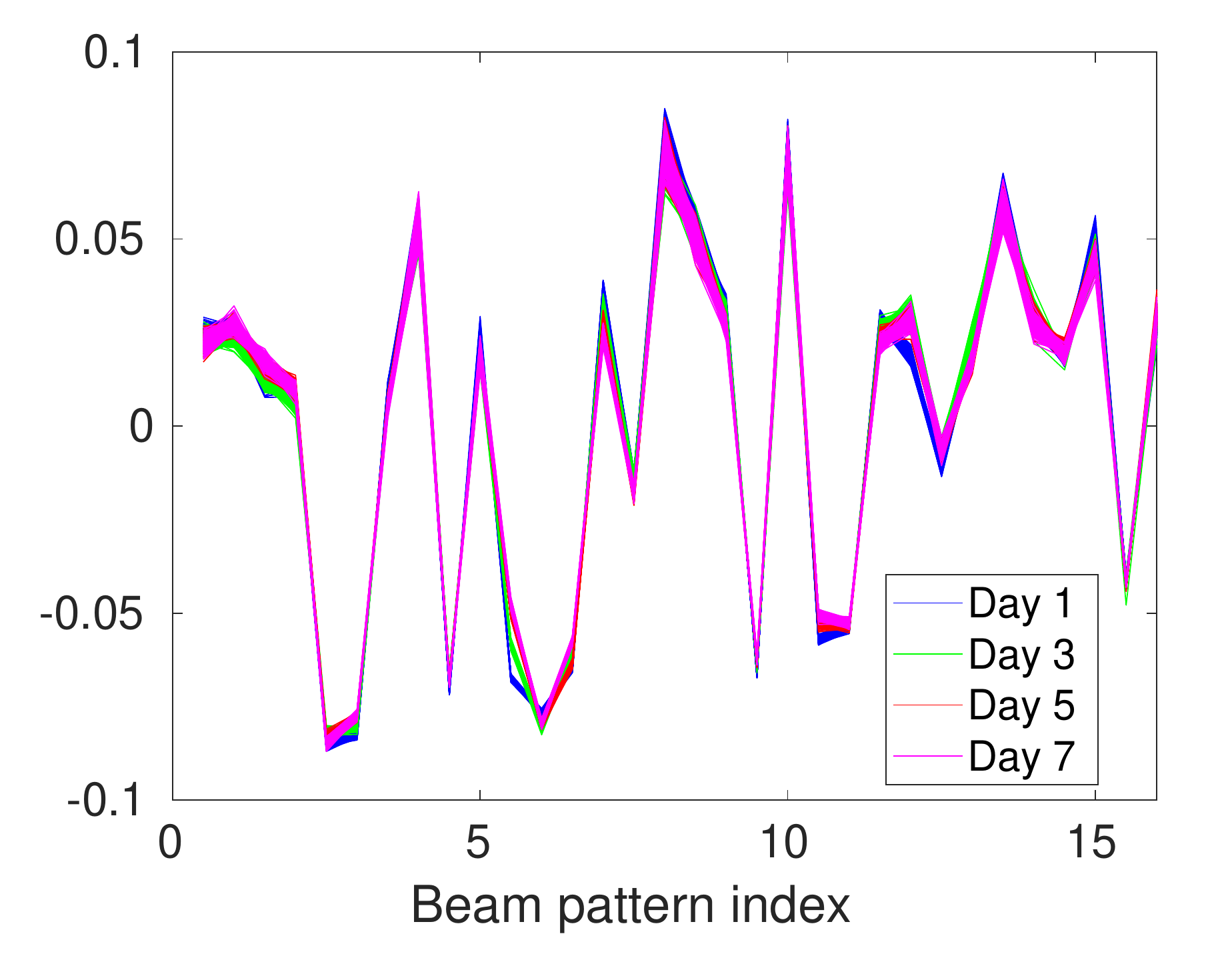}
  \caption{Stability of fingerprints.}
  \label{fig:stability}
  \end{minipage}
\end{figure}
Another important characteristic of the suitability of a fingerprint for device identification is stability\cite{danev2012physical}. The obtained fingerprint should be invariant over time. To verify the stability of our proposed beam pattern feature, we set up a Talon AD7200 router as client and recorded 200 beam searching beacons everyday from it over a period of 1 week. Fig. \ref{fig:stability} shows that the extracted beam pattern feature is stable over a long period of time.   

\paragraph*{\textbf{User Capacity}}
\label{sec:capacity}
Previously, we discussed uniqueness of the beam pattern feature among different antenna arrays with identical geometries and materials used. However a natural question arises: \textit{how many such devices (antenna arrays) can have unique beam patterns?} If the number of devices enrolled in the system exceeds the capacity of the system, then the beam pattern features of devices will overlap decreasing the accuracy of the identification system. Therefore, it is important to study the user capacity of the proposed beam pattern feature based on the characteristics of the antenna array and the limitations of the fingerprinting device. As discussed previously, the beam pattern of the antenna array significantly deviates from the theoretical beam pattern due to the errors introduced by the tolerances associated with materials used and the fabrication process. With the antenna array in Sec. \ref{sec:array design}, we vary the substrate dielectric ($5.9 \pm 0.2$), dimensions of the patch (X,Y shrinkage $15\% \pm 2$), and phase ($\theta_{i} \pm 5^{\circ}$) of the phase shifter within their respective tolerance values and perform parametric sweep simulation using COMSOL. The resulting beam patterns for various combinations of the parameters is shown in Fig. \ref{fig:comsol_beampattern}. Here we are interested in finding the variance of the beam pattern $f_{\theta}$ due to the errors. Also it should be noted, since beam pattern is a function of $\theta$, the maximum deviation of the beam pattern is also a function of $\theta$. As discussed in Sec. \ref{sec:learning procedure}, we learn multiple angular directions $\theta$ of the beam pattern $f_{\theta}$ of the user. We find that, for the antenna array in Sec. \ref{sec:array design}, the maximum variation of the beam pattern $f_\theta$ for all $\theta$ to be $3.2dB$. 
\par Recall that, the beampattern of the device $f_{\theta}$ is obtained at the receiver by measuring the signal power at $\theta$. Hence the resolution of the digitizer used at the receiver plays an important role in further determining the user capacity of the system. The analog-to-digital converter (ADC) has a maximum voltage $V_{max}$ and minimum voltage $V_{min}$ it can sample without distortion with full scale $V_{FS}=V_{max}-V_{min}$. The number of discrete voltage levels the ADC can output depends on the number of bits $n$ it uses to represent a voltage level. For a $n$ bits ADC, the voltage resolution $\Delta v$ is given by $\frac{V_{FS}}{2^n-1}$. Taking into account the voltage resolution of a 13-bit ADC, for a beam pattern the maximum number of users the fingerprinting system can support is 3200. Since our fingerprinting system uses all the codebooks transmitted during beam searching phase, the user capacity scales with the number of beam patterns used. For a mmWave device with $N$ codebooks, the maximum user capacity is $3200 \times N$.

\subsection{AP Deployment}
\label{sec:AP deployment}

\begin{figure}[htbp!]
\subfloat[\label{fig:floor plan}]{\includegraphics[width=0.24\textwidth]{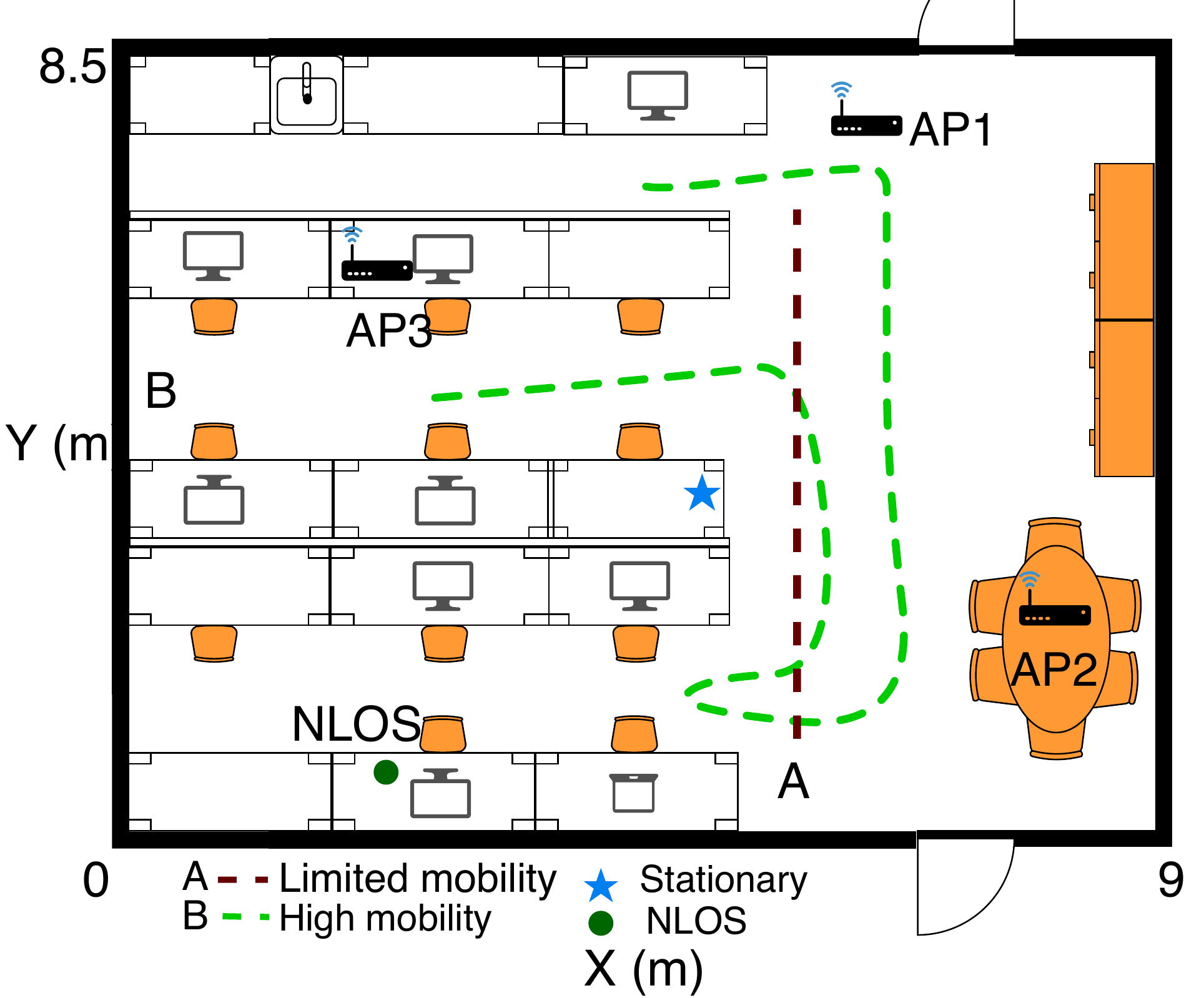}}
\subfloat[\label{fig:AP locs}]{\includegraphics[trim={2cm 6cm 2cm 7cm},clip,width=0.24\textwidth]{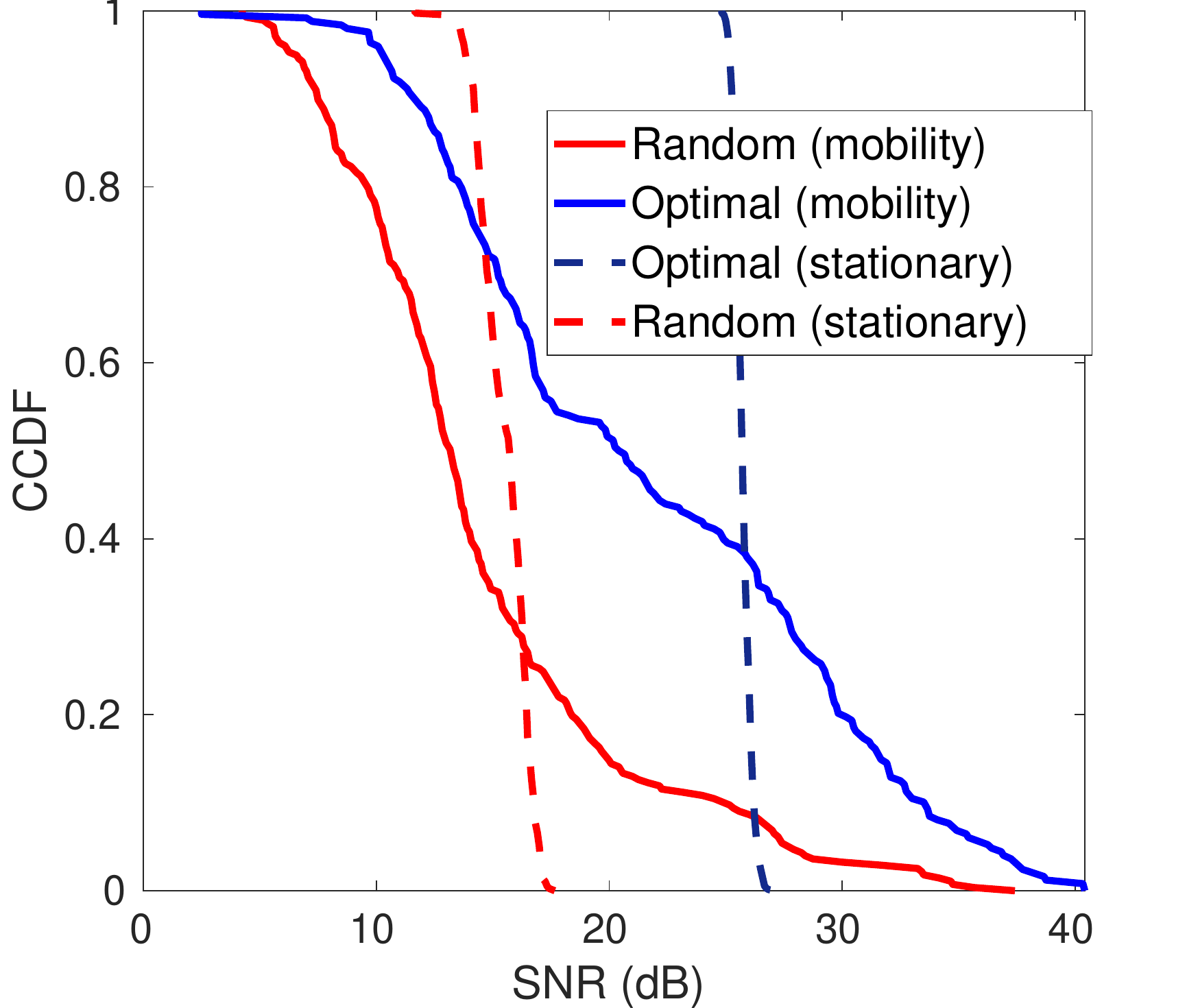}}
\caption{(a) Floor plan of the experimental area and (b) distribution of average signal strength variations at 3 APs for random deployment strategy and optimal deployment strategy (\ref{eq:AP location}) under stationary and device mobility.}
\end{figure}
As discussed in the previous sections, the mmWave beam searching mechanism allows the APs to obtain high dimensional beam pattern features from the devices. However, from a particular spatial position, each AP can only obtain a single angular view of each beam pattern used by the client. Deploying multiple APs in the area will allow the APs to have multiple views of the beam pattern of the client. But if the APs are deployed too close to each other or deployed in a position with obstacles, the beam pattern feature obtained by them might not offer distinct information. Therefore the deployment position of APs in our proposed mmWave device identification system plays an important role. The APs have to be deployed such that the device to be identified is in the signal coverage area. In this section, we discuss a practical deployment strategy for the APs.
\paragraph*{\textbf{Optimal Deployment}}
\par We propose a method to optimally deploy APs which can be applied to any environment and does not require extensive measurement campaigns. The APs are to be deployed in an area $U$. For a certain user position $P_u$, the probability the user is in the coverage area of the AP, i.e., the beacons from the user reach the AP can be expressed as, $P_{cov}=\sum_{i}P(P_{u} \text{is in} A_i).P(SNR(P_{u})>T|P_{u} \text{is in} A_i)$, where $A_i$ is the coverage area of AP $i$ and $T$ is the threshold for the signal to be received. $SNR(P_{u})$ is the SNR at $P_{u}$. The user location $P_u$ is assumed to follow a certain probability distribution with density $f(P_u)$. We assume the user's movement follows a mobility model as in \cite{ge2016user}. For the user mobility, the probability the user visits a new position is given by $P_{new}(n)=\alpha S^{-\gamma}(n)$ and the probability of visiting an old position is given by $P_{old}(n)=1-P_{new}(n)$ where $0 < \alpha < 1$ and $\gamma > 0$. $S(n)$ is the number of visited positions in $n$ number of jumps. The probability the user departs the area covered by an AP is given by,
$P_{d} = \frac{1-\Bigg\{\frac{1}{\bar{v}}\bigg[(\frac{A}{U})^2\mathbb{E}[d_{i,i}]+(1-\frac{A}{U})\frac{A}{U}\mathbb{E}[d_{o,i}]\bigg]+\frac{A}{U}\mathbb{E}[\Delta t_{i}]\Bigg\}}{\Bigg\{\frac{1}{\bar{v}}\bigg[(\frac{A}{U})^2\mathbb{E}[d_{i,i}]+2(1-\frac{A}{U})\frac{A}{U}\mathbb{E}[d_{o,i}]\bigg]+(1-\frac{A}{U})^2]+\frac{A}{U}\mathbb{E}[\Delta t_{i}+(1-\frac{A}{U})\mathbb{E}[\Delta t_{o}]\Bigg\}}$
where $\Delta t_i$ is the user's waiting time inside the area covered by the AP and its distribution is given by $f(\Delta t_i)=|\Delta t_i|^{-1-\beta_i}$ with $0< \beta_i \leq 1$. $d_{i,i}$ is the users jump distance within the coverage area and $d_{o,o}$ is the distance between two points outside the coverage area. $d_{o,i}$ is the distance between one point outside and one point inside the coverage area. $\bar{v}$ is the average velocity of user mobility. Therefore the probability the user and the AP are able to successfully receive each other's signal is given by
$P_{cov,mobility} = (1-P_d)P_{cov}.$  
For multiple APs, the connection probability for the $i$th AP is given by $P_{cov}=\int_{A_i}F(SNR(d_{i}))f(P_{u})ds$ where $A_i$ is the area covered by the $i$th AP, $d_i$ is the distance between the user and the $i$th AP given by $d_i=||P_{AP,i}-P_{u}||$. The connection probability can be further written as 
$P_{cov} = \sum_{i}\int_{A_i-\cup_{j\neq i}A_{j}}F\left (SNR(d_{i}))\right)f(P_{u})ds 
+\sum_{n=1}^{N_{o}}\int_{A_{o}}{1-\prod_{i}\Bigg[1-F\left(SNR(d_{i})\right)\Bigg]}f(P_{u})ds$,
where $A_{o}$ is the area of the overlapped coverage region due to multiple APs and $N_{o}$ is the number of overlapped regions. $F(x)$ is an indicator function with $F(x)=1$ for $x \geq T$ and $F(x)=0$ for $x<T$. 
\par Therefore for the user and AP to successfully receive each other's beacon signals, the APs can be deployed so as to maximize the signal coverage probability. I.e.,
\begin{equation}
max_{P_{AP,i}} P_{cov,mobility}
\label{eq:AP location}
\end{equation}
The optimal positions of the APs for the deployment area shown in Fig.\ref{fig:floor plan} is found by solving (\ref{eq:AP location}). To reduce the complexity in searching for the optimal AP positions $P_{AP,i}$, we can solve (\ref{eq:AP location}) for predefined user supplied AP positions known through environment familiarity or divide the deployment area into grids and provide the center of the grid locations as user supplied AP positions. We used the grid approach to find the AP positions that maximizes (\ref{eq:AP location}).
\paragraph*{\textbf{Experiment Validation}}
We compare the AP deployment solution from (\ref{eq:AP location}) with a random deployment strategy. For the optimal strategy, the APs are placed at locations returned by (\ref{eq:AP location}) and for random deployment, the APs are positioned at 3 random locations in the room. The SNR of the client to each of the AP is measured at the AP and averaged across all the APs. Fig. \ref{fig:AP locs} plots the CCDF of signal coverage for different client mobility scenarios: a) stationary, and b) mobility under random deployment strategy and optimal deployment strategy. We set the minimum SNR required to receive the beam searching beacon to 12 dB (corresponding to -78dBm control signal threshold as specified in 802.11ad standard and -90dBm noise floor). We see that, both the random and the optimal strategies exceed the SNR requirement, however the optimal strategy exceeds 25dB SNR whereas random deployment could only achieve 15dB SNR $90\%$ of the time. For the mobility experiment, we see that the optimal strategy exceeds the SNR requirement for $95\%$ of the time whereas the random deployment exceeds the SNR requirement only $75\%$ of the time. The signal quality gained through environment aware optimal AP deployment strategy directly translates to improved identification accuracy for the devices. 

\subsection{Feature Extraction Protocol}
\label{sec:learning procedure}
\begin{figure}[htbp!]
\begin{minipage}[b]{0.6\columnwidth}
\includegraphics[trim={1cm 0 11cm 0},clip,height=4.5cm,width=\columnwidth]{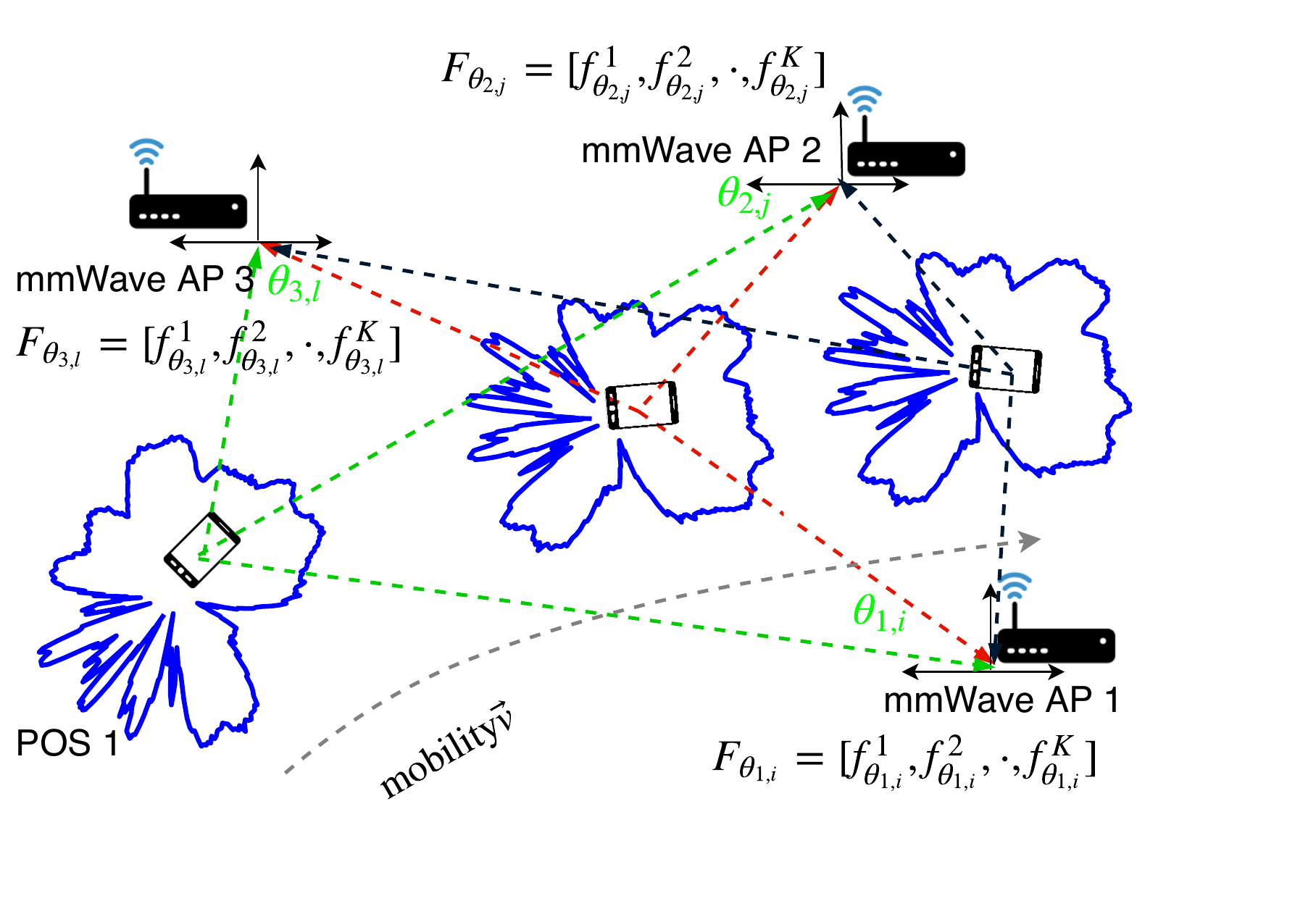}
\caption{Beam pattern feature extraction using multiple APs.}
\label{fig:feature_extraction}
\end{minipage}
\begin{minipage}[b]{0.38\columnwidth}
\subfloat[Stationary]{\label{fig:stationary_codebook}\includegraphics[trim={0cm 6cm 1cm 6cm},clip,scale=0.15]{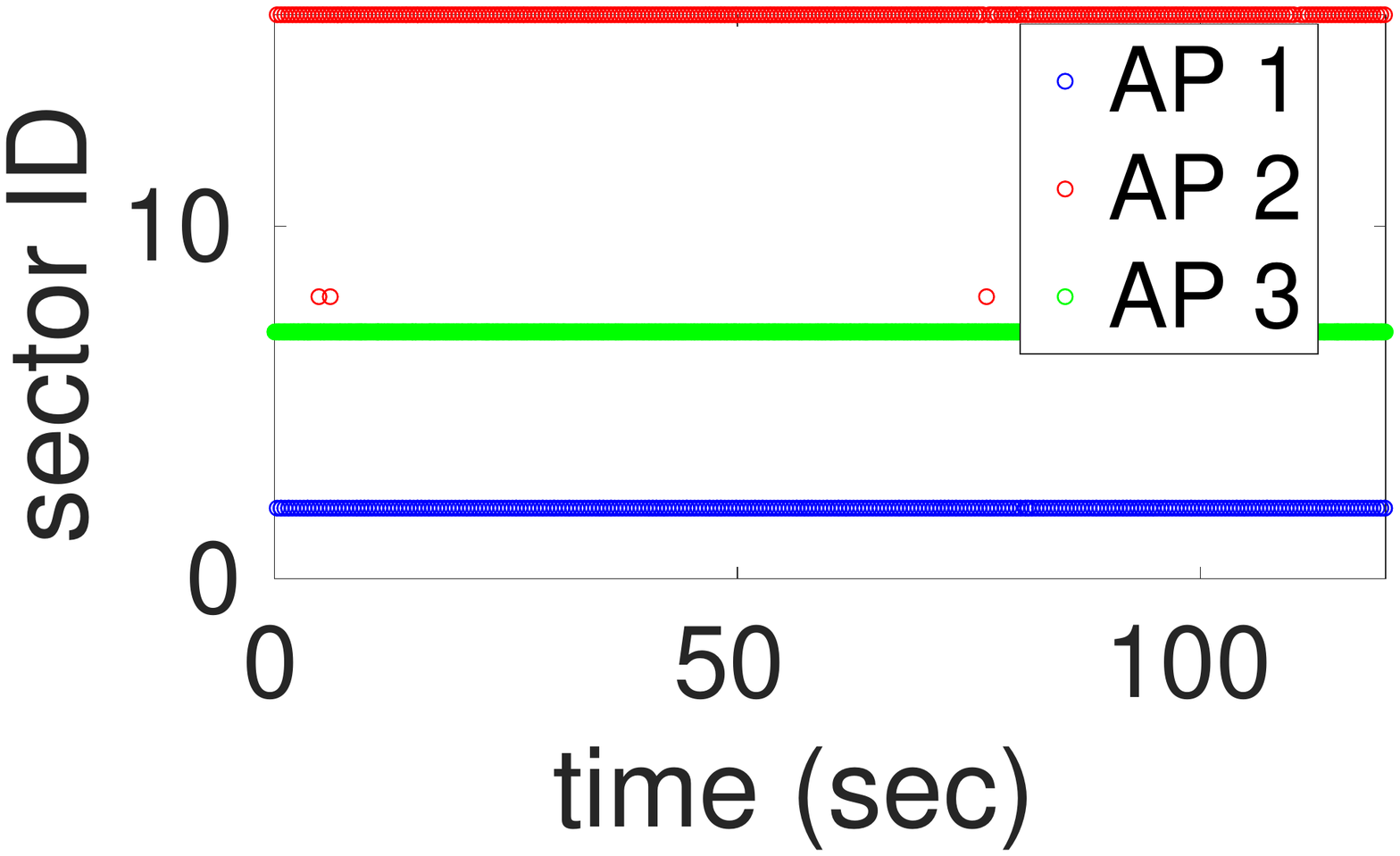}} \vspace{0.05em}
\subfloat[Mobility]{\label{fig:mobility_codebook}
\includegraphics[trim={0cm 6cm 1cm 6cm},clip,scale=0.15]{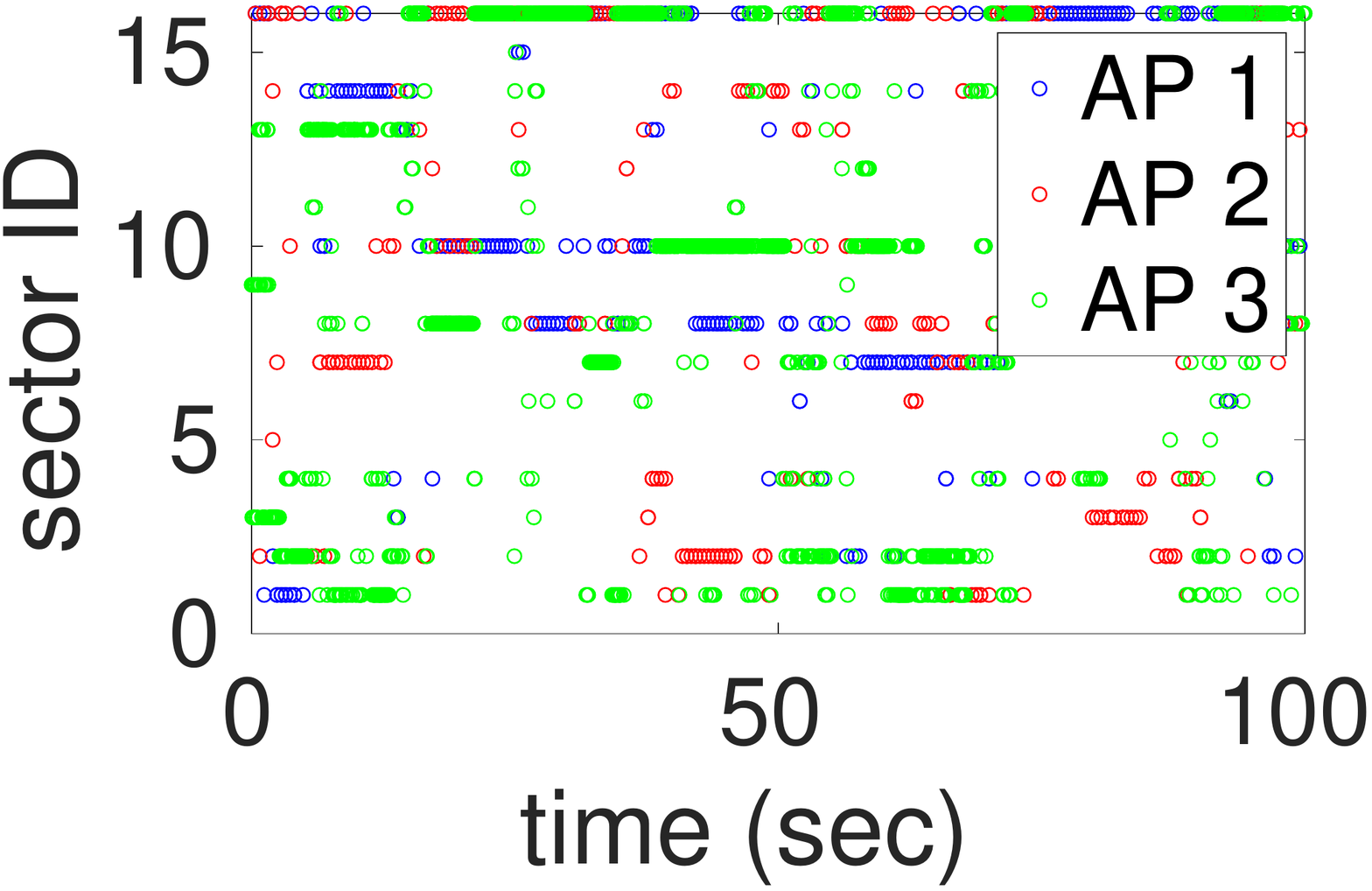}}
\caption{Beam pattern view at 3 APs.}
\end{minipage}
\end{figure}
\par Now, having discussed the source of our proposed beam pattern feature, its uniqueness and potential applicability to mmWave device fingerprinting, \textit{how do we measure the beam pattern in practical scenarios?} MmWave devices transmit beacons through each of their sectors with unique beam patterns. For e.g., the Talon AD7200 router transmits 32 different beam patterns in each of its 32 sectors during beam searching, a process known as sector level sweep (SLS) in 802.11ad standard. We can utilize the SLS or beam searching process to measure the beam pattern $f^{k}_{\theta}$ of the $k$th beam pattern. During the beam searching process, the user device either initiates beam searching with the APs or responds to beam searching beacons from the APs, during which the user transmits a beam pattern $k \in K$ in one of its sectors. AP 1 measures $f^{k}_{\theta_{1,i}}$ (APs measure the received signal strength of beam patterns, a process inherent to 802.11ad/ay and 5G-NR standards), assuming $\theta_{1,i}$ is the orientation angle of AP 1 with respect to the user in global coordinate system. Now, as the user transmits every beam pattern $k \in K$, AP 1 measures $f^{k}_{\theta_1,i}$ for each beam pattern $k$ and builds the feature vector $F_{\theta_1,i}=[f_{\theta_1,i}^{1},f_{\theta_1,i}^{2},\cdots,f_{\theta_1,i}^{K}]$. Each element $f_{\theta_{1,i}}^{k}$ in the vector $F_{\theta_{1,i}}$ is contributed by a distinct angle $\theta_{1,i}$ in one of the $k \in K$ beam patterns used by the user device during beam searching. When multiple APs are available as shown in Fig. \ref{fig:feature_extraction}, each AP $m$ will have distinct view $\theta_{m,i}$ of the $k$th beam pattern, where $m=1,\cdots,M$ with $M$ being the number of APs, to form the vector $F_{\theta_{m,i}}=[f_{\theta_{m,i}}^{1},f_{\theta_{m,i}}^{2},\cdots,f_{\theta_{m,i}}^{K}], m=1,...,M$. Recall that, each AP only has a distinct view $\theta$ of a particular beam pattern $k$. We illustrate this scenario where the devices are stationary with respect to the APs. 
\paragraph*{\textbf{Stationary devices}}
For a stationary scenario with fixed device orientation angle relative to the APs, the beam pattern vector $F_{\theta_{m,i}}$ of the codebooks swept by the device during the beam searching mechanism does not change with respect to the device-AP direction. Fig. \ref{fig:stationary_codebook} show such a scenario where the orientation in terms of best sector ID from the user to AP direction remains same for 500 beam searching periods spanning $120 secs$. AP 1, AP 2 and AP 3 always see best sector 2, 7, and 16 respectively from the user device. The APs could simply extract the beam pattern vector $F_{\theta_{m,i}}$ of the $K$ codebooks over several beam searching periods and build the fingerprint database for that device. However, if the device moves or the orientation of the device changes with respect to the APs, then the beam pattern fingerprint database needs to be updated as the beam pattern feature vector $F_{\theta_{m,i}}$ changes with change in orientation of the device with respect to the AP. 
To authenticate a mobile user, the beam pattern $f_{\theta}^{k}$ of the $k$th codebook for all feasible $\theta$ must be measured.
\par \textit{Now how do we learn all possible views of a particular codebook $k$?} Next, we discuss mobility scenario under which multiple views of a codebook $k$ could be learned.

\paragraph*{\textbf{Beam pattern feature due to mobility}} 
MmWave devices initiate the beam searching process whenever the best beam found during the previous beam searching phase becomes outdated due to device mobility or orientation change. When the device moves or orientation changes, the angle $\theta_i$ of the $k$th codebook of the user device seen by the AP changes as shown in Fig. \ref{fig:feature_extraction}. A simple rotation of the device by $d$ degrees from initial orientation $\theta$ will result in beam pattern $f_{(\theta + d)}$ at the AP. \textit{Can we use device mobility to learn the beam pattern of the codebooks used by the devices?} To answer this, we performed experiments with APs deployed in the optimal position (Sec. \ref{sec:AP deployment}) in the environment and orientation angle in terms of best sector ID of the user with respect to each of the AP is measured for $100s$ of random user mobility. We see from Fig. \ref{fig:mobility_codebook}, the best sector ID of the user changes with respect to each of the three APs. The APs learn distinct beam pattern points $f_{\theta_{m,i}}^{k}$ for each of the $k$ codebooks everytime the user moves or device orientation changes. 

\subsection{Mobility vs. Multiple APs}
We discussed in Sec. \ref{sec:AP deployment} and Sec. \ref{sec:learning procedure} how employing multiple APs and the user device mobility helps in learning the rich spatial features of the beam patterns used by the devices. Here we discuss the trade-off among multiple APs, mobility and security implications. \begin{itemize}

\item In a single AP system, the AP can only use one distinct view of the user beam pattern to authenticate it which makes the system vulnerable to impersonation attack as shown in Sec. \ref{sec:AntennaPatternAttack}. Employing multiple APs would allow the APs to utilize multiple views of the user beam pattern to authenticate it which drastically improves the security of the system against impersonation attacks. However, multiple APs increases the infrastructure cost of the system.
\item One way to reduce the cost and increase the security is to take advantage of device mobility. If the user is mobile, a single AP is sufficient and the AP could wait until it learns more distinct views in a beam pattern used by the user and proceed to authenticate the clients. In this way, it will be hard for the attacker to know the number of views used by the AP for authentication and spoof it. 
\end{itemize}

\subsection{Classification Architecture} \label{sec:cnn}
Given a set of beam pattern features and associated target device IDs for each of those features, the objective of the classifcation problem is to learn a function or model that uniquely maps the beam pattern feature to its originating device. We model the mmWave device identification task as a multi-class classification problem. The input to the classifier is the beam patterns extracted during the beam searching process. The outputs are corresponding class labels. We propose to use a 4 layer convolutional neural network (CNN) architecture for our classification system. The network consists of two 1D convolutional layers and two dense layer with ReLu activation function and a fully connected softmax layer. The convolutional layers use a filter size of 256 and 80 respectively and are initialized with Glorot uniform initializer. The dense layers were initialized with He normal initializer. Categorical cross entropy loss function and Adam solver is used for training. The input feature dimension to the network is $N \times M \times K$ where $N$ is the number of training samples, $M$ is the number of APs and $K$ is the feature length.
To reduce the effect of amplitude variations due to user-AP distance, the input beam pattern feature vector is z-score normalized with $\hat{F}_{\theta_{m,i}}=\frac{F_{\theta_{m,i}-mean(F_{\theta_{m,i}})}}{std(F_{\theta_{m,i}})}$. 
To train the network, the fingerprint database is randomly split into training set and validation set. The training is validated using the validation set for each epoch of training. An early stopping criterion is utilized to stop the training when the validation loss does not minimize after 5 number of evaluations of the validation set. The CNN is implemented in Python using Keras \cite{chollet2015keras} and trained on an Intel Core i7-5500U machine.

\section{System Implementation And Test-bed Development}
\label{sec:system implementation}
\begin{figure}
\centering
\includegraphics[trim={3cm 3.5cm 9cm 1cm},clip,width=0.3\textwidth]{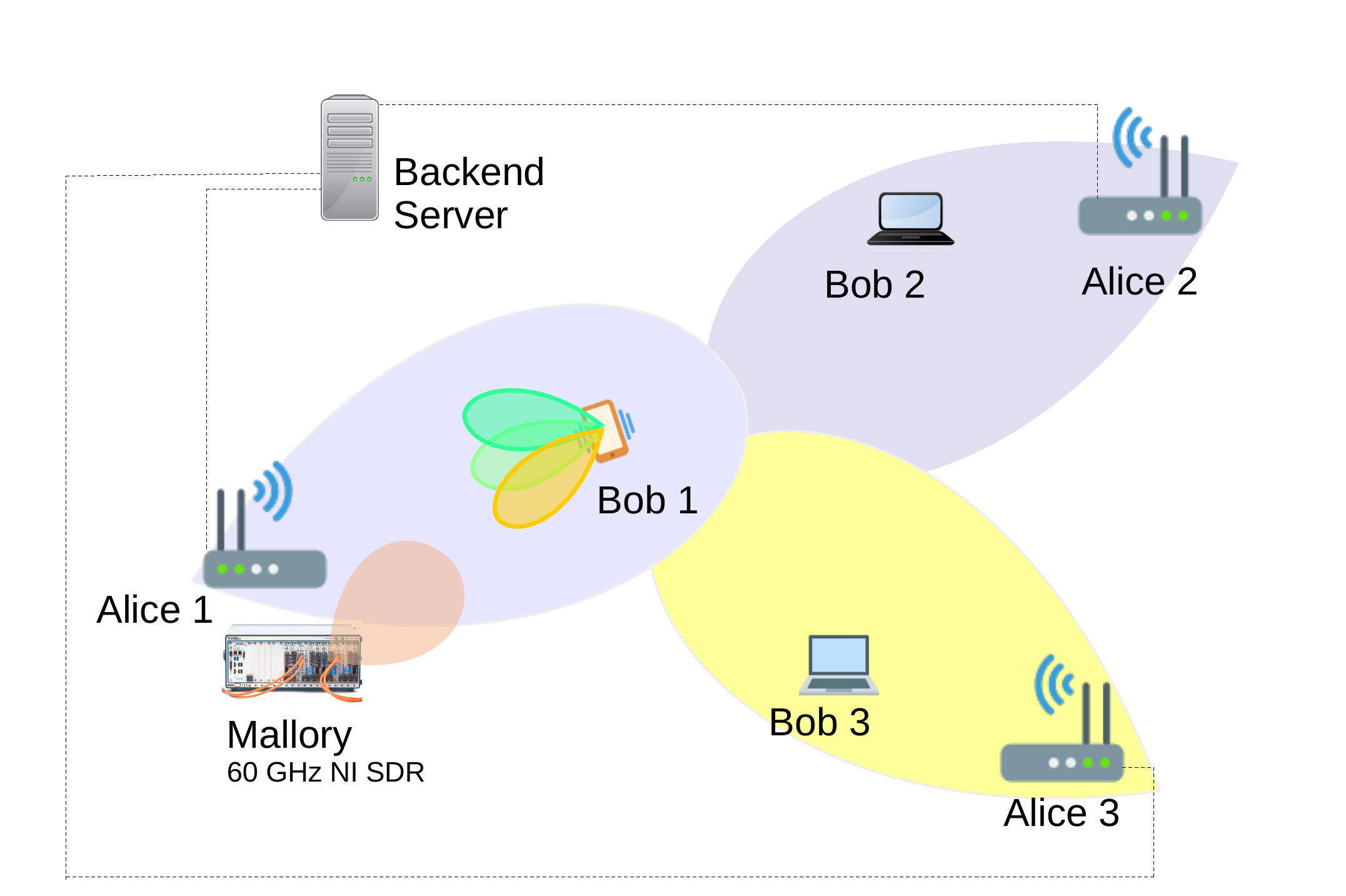}
\caption{60 GHz wireless network RF fingerprinting: Scenario showing multiple-APs (Alice's) and Bob's in the wireless network. Attacker \textit{Mallory} launches impersonation attack on legitimate user \textit{Bob 1} using state-of-the-art 60 GHz SDR. }
\label{fig:IdentificationAndAttackScenario}
\end{figure}
In this section, we describe our system implementation and a testbed we set up for mmWave RF fingerprinting. Fig. \ref{fig:IdentificationAndAttackScenario} shows a real-world application scenario where a mmWave wireless network is deployed with multiple APs and multiple users connected to the wireless network. For the rest of the discussion we refer to the APs as \textit{Alice 1}, \textit{Alice 2}, and so on and users as \textit{Bob 1}, \textit{Bob 2}, and so on. Without loss of generality, we assume that \textit{Bob 1} is connected to \textit{Alice 1} after successful completion of authentication procedure using the proposed RF fingerprinting. Mallory $\mathcal{M}$ is a malicious node that wants to gain unauthorised access to the network by forging the identity of legitimate node \textit{Bob 1}. For the scenario described, first we discuss the reliability of the proposed beam pattern fingerprint in enabling Alices to authenticate legitmate Bobs under various scenarios. Next, we analyse the security impact of the system by evaluating the ability of Mallory to impersonate the identity of legitimate node \textit{Bob 1} by performing signal replay attack.        

\par 
Before proceeding to discuss the reliability and security analysis, we detail the mmWave devices, experiment setting and signal acquistion method in detail.
\subsection{\textbf{mmWave devices}} 
Here we present the mmWave devices used for reliability and security analysis experiments.
\paragraph{Alice and Bob} We have evaluated the reliability and security of our proposed fingerprinting method discussed in Sec. \ref{sec:system design} on commercially available mmWave devices. We conduct experiments on the following devices: 3 TP-Link Talon AD7200 routers, 3 Netgear Nighthawk X10 Smart WiFi routers, 1 Acer Travelmate P446M laptop with Qualcomm 60 GHz NIC and 6 Intel Tri-band 18625 NICs equipped with antenna model antenna-M 10101-R. 
The devices are labeled as follows: Talon1, Talon2, Talon3, Netgear1, Netgear2, Acer1, Intel1, Intel2, Intel3, Intel4, Intel5, Intel6. All these devices follow the 802.11ad standard and perform beam searching procedure as outlined in \cite{80211adstandard}. We use 802.11ad based devices since at present they are the only mmWave devices that are available in the market. Nevertheless, the proposed method and the findings in our experiments are applicable to other mmWave wireless standards such as IEEE 802.11ay and 5G-NR as they also utilize antenna array and beam searching mechanism to establish directional communication.
\paragraph{Attacker Mallory} 
\label{sec:Attacker device}
The attacker $\mathcal{M}$ uses one X60 node \cite{saha2018x60}, to record the transmissions from legitimate device and retransmit them. X60 nodes are based on the 60 GHz software defined radio. The X60 node has a Si-Beam 60 GHz antenna array. The antenna array has 12 transmit elements and 12 receive elements with reconfigurable codebooks. The codebooks can be configured through setting the phase of each element of the antenna array. In our experiments, we use codebook 12 with beam pattern shown in Fig. \ref{fig:NI beam pattern}. Before performing the attack, the RF chain in the transmitter is calibrated. The RF chain introduces I/Q imbalance and DC offset to the transmitted signal. Also the synthesizer in the transmitter introduces LO leakage which can degrade the quality of the replayed signal.
To process the signal from the ADC as well as to transmit baseband signals through the X60 testbed, a FPGA module is designed using Labview. The FPGA module interfaces with the ADC  of the X60 node while in the receiver mode and interfaces to the DAC of the X60 node while in the transmitter mode. 
\subsection{\textbf{Signal Acquisition}} 
Since the signal strength measurements made during the beam searching process is not available outside of the firmware for most of the devices other than TP-Link Talon AD7200 \cite{talon-tools:project}, therefore to maintain uniformity, we receive  the beam patterns swept during beam searching phase using VubiQ 60GHz mmWave receiver. We noticed that the mmWave devices we used in our experiments use channel 2 (60.48 GHz) of the 60 GHz band to perform beam searching.  The VubiQ receiver is tuned to 60.48 GHz and it downconverts the received signal to analog baseband. The analog baseband signal is sampled using an Agilent oscilloscope. The oscilloscope is controlled by a Matlab script to acquire the signal and save them in the hard drive.  

\section{Reliability Analysis And Experimental Evaluation}
\label{sec:experiments}
We implement the proposed beam pattern fingerprinting system and evaluate its effectiveness in various settings. We compare our beam pattern fingerprint with another conventional fingerprint method namely PSD fingerprint. To that end, we implement PSD feature based fingerprinting for mmWave devices. Before proceeding to evaluate our beam pattern feature, we first present the PSD feature extraction method.
\subsubsection{Spectral Feature}
\label{sec:psd}
We use PSD of the received signal as a feature for device classification and identification. The PSD is a frequency domain feature and is estimated at the baseband of the received signal. Let $y(n)$ be the received signal. The normalized spectral fingerprint $S(n)$ is given by
\begin{equation}
S(n)=\frac{|Y(n)|^2}{\sum_{n}|Y(n)|^2}.
\label{eq:psd}
\end{equation}
$Y(n)$ is the $N$ length Discrete Fourier Transform of the received signal $y(n)$. The length of the resulting feature vector depends on the DFT length $N$. The effect of DFT length $N$ on the classification performance is discussed in Sec. \ref{sec:training size}. 

\paragraph*{Feature Extraction and Classification Method}
\begin{figure}[!htbp]
  \centering
  \subfloat[\label{fig:preamble_corr}]{\includegraphics[width=0.24\textwidth]{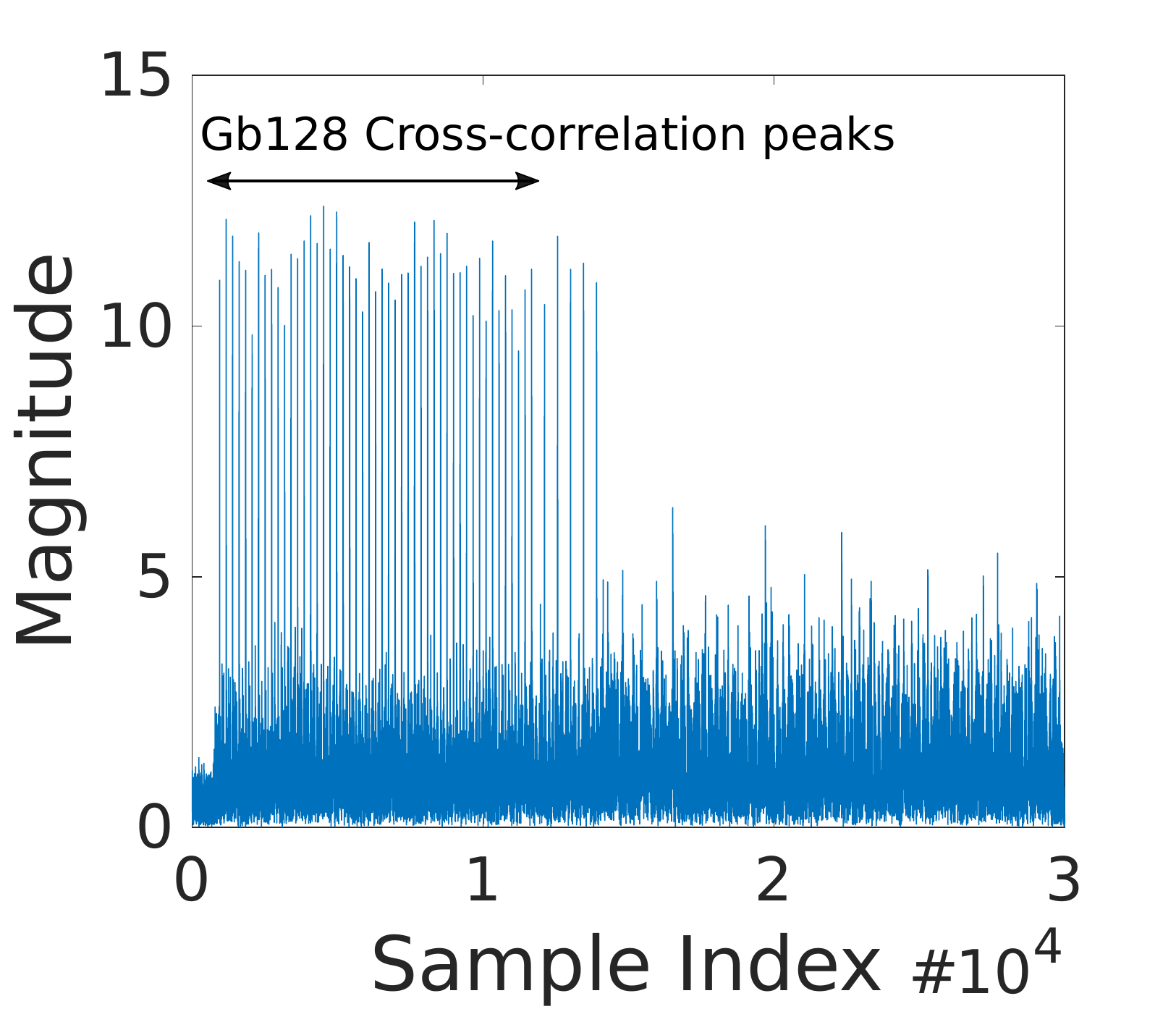}}
  \subfloat[\label{fig:psd}]{\includegraphics[trim={2.5cm 6cm 2cm 7cm},clip,width=0.24\textwidth]{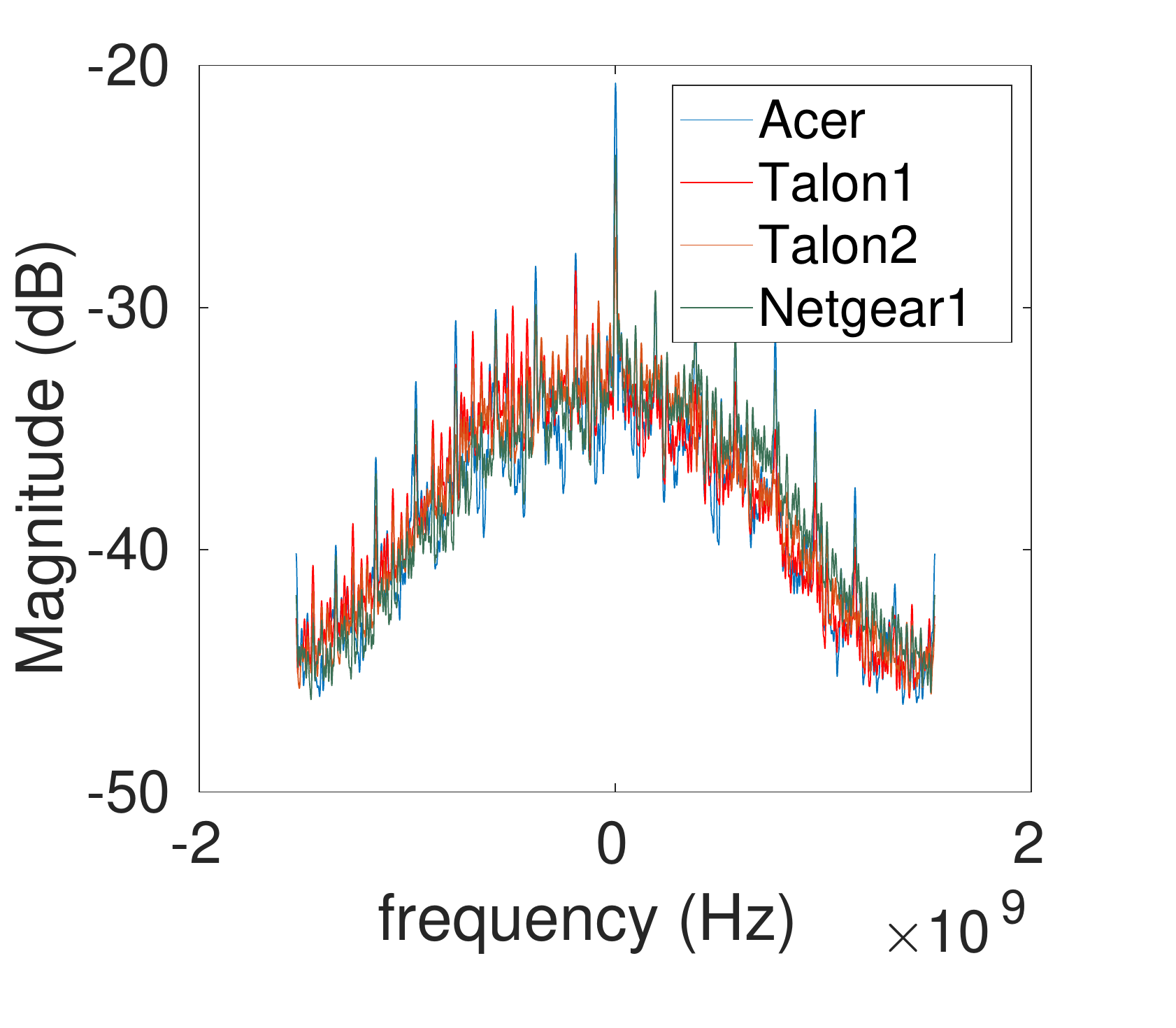}}
  \caption{(a) Correlation of the received beacons with $Gb_{128}$ sequence and (b) PSD feature of devices.}
\end{figure}

The PSD is obtained from the preamble part of the received signal. We use beacons transmitted by the mmWave devices during the beam searching process to extract the PSD feature. The proposed method is applicable to all mmWave wireless standards and in this section, we focus on 802.11ad standard since we use 802.11ad devices for our experiments. 802.11ad devices transmit the control PHY preamble \cite{80211adstandard} during the beam searching process. The preamble part of the beacon transmitted during beam searching consists of short training (STF) field and channel estimation (CE) field. The STF consists of 48 repetitions of length 128 Golay sequence $Gb_{128}$ followed by a single $-Gb_{128}$ Golay sequence and a $-Ga_{128}$ Golay sequence. We use the STF and CE part of the preamble for PSD feature extraction. The beginning of the beacon frame is detected by the normalized auto-correlation of the known STF sequence with the received beacon frame \cite{kumari2018ieee}. Fig. \ref{fig:preamble_corr} shows the preamble correlation to detect the beginning of the beacon frame. Once the frame is detected, the STF and CE part of the frame are used for PSD estimation using (\ref{eq:psd}). The estimated PSD feature for three of the devices is shown in Fig. \ref{fig:psd}. The feature vector length is $1 \times NFFT$ where NFFT is the number of FFT points used to compute PSD. The classification method described in section \ref{sec:cnn} is used for PSD feature classification and identification with the input dimension of the convolutional neural network set to $1 \times NFFT$.
\par Next we proceed to discuss the reliability of our proposed beam pattern fingerprint and compare with the PSD fingerprint in terms of classification and identification accuracy. 

\subsection{{Experimental setting and metrics}} For the reliability analysis experiments, we broadly classify our experimental setting into the following scenarios: 1) stationary LOS, 2) stationary non line-of-sight (NLOS), 3) limited mobility and 4) high mobility. All the experiments are performed in the indoor laboratory area with floor plan shown in Fig. \ref{fig:floor plan}. For all the experiments, the locations of the APs are obtained from Sec. \ref{sec:AP deployment} and fixed. They are denoted as AP1, AP2 and AP3 in Fig. \ref{fig:floor plan}. To evaluate our proposed beam pattern feature, we present the following metrics: 1) average accuracy metric which is the ratio of correctly predicted observations to the total predicted observations, 2) Receiver Operating Characteristic (ROC) curve which is obtained by plotting the true positive rate (TPR) against the false positive rate (FPR) for various thresholds and 3) Equal Error Rate (EER), an operating point in ROC at which false accept rate (FAR) = false reject rate (FRR). Lower the EER value, better is the system performance.

\subsection{Classification Results}
\paragraph*{\textbf{Training size and classification accuracy}}
\label{sec:training size}
\begin{figure}[!htbp]
\begin{minipage}[b]{0.48\columnwidth}
  \includegraphics[trim={2cm 6.8cm 3cm 7.5cm},clip,width=\columnwidth]{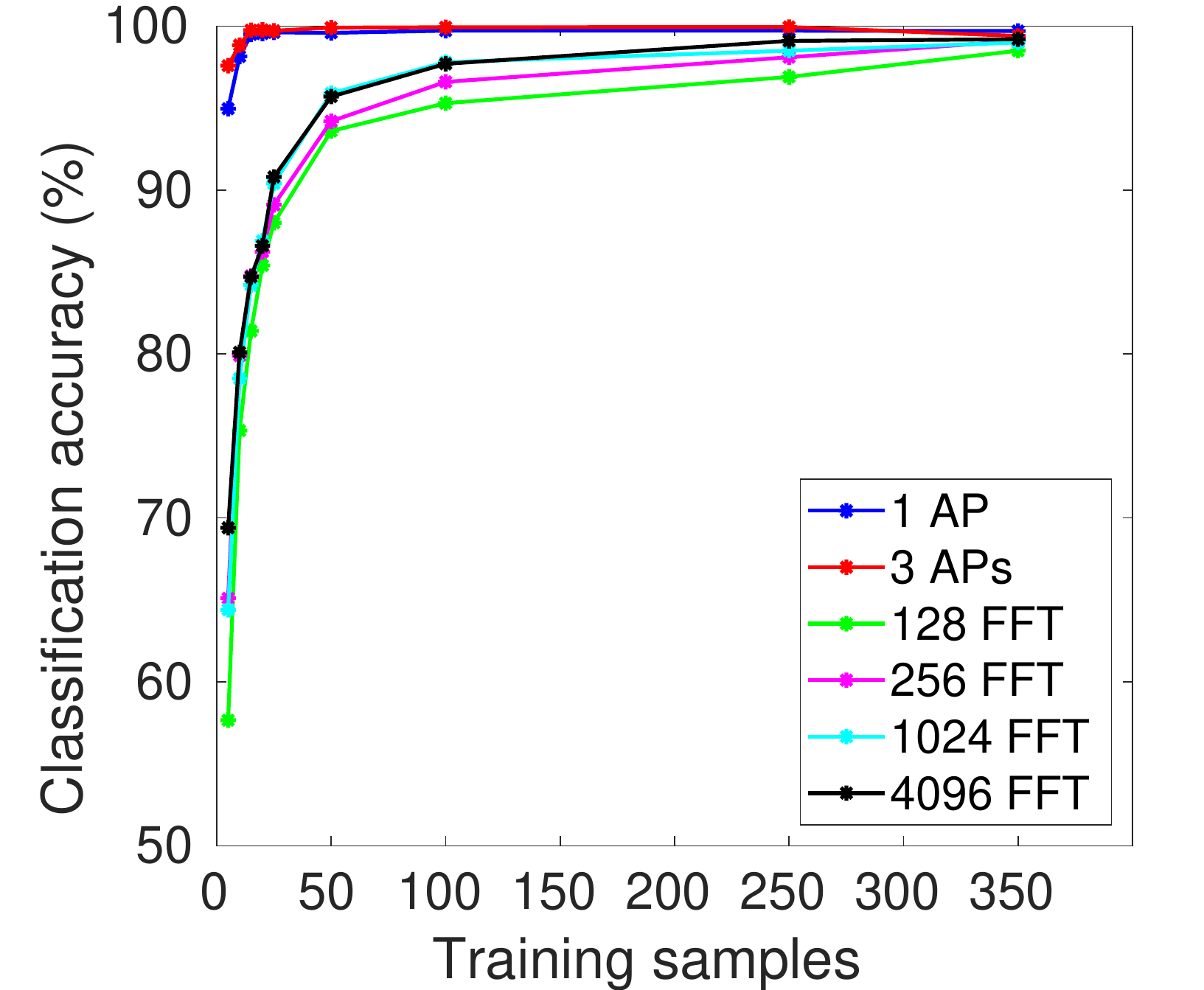}
    \caption{Classification accuracy for different training sample size. }
  \label{fig:accuracy_samples}
\end{minipage}
\begin{minipage}[b]{0.48\columnwidth}
  \includegraphics[trim={0 0 0 3cm},clip,width=\columnwidth]{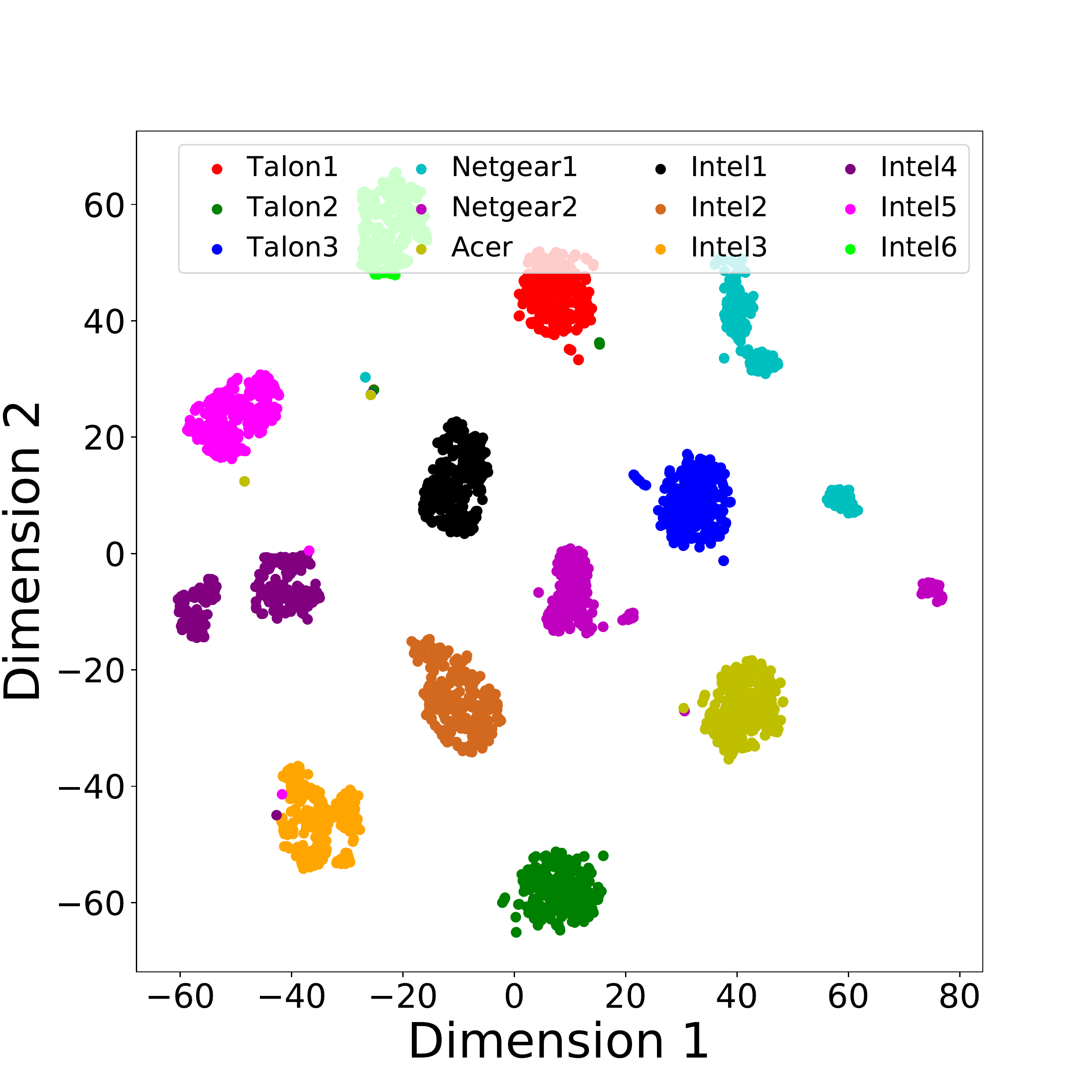} 
  \caption{2-Dimensional embedding of the feature learned by CNN.}
  \label{fig:classifier_3AP_tsne}
  \end{minipage}
\end{figure}
In this section we experimentally evaluate the number of training feature vectors needed to train the classifier in Sec. \ref{sec:cnn} to achieve a certain overall accuracy and in doing so we seek to answer the following question: \textit{How well is the classifier in Sec. \ref{sec:cnn} able to distinguish between devices of same manufacturer and across manufacturers?} We vary the training size for each evaluation while keeping the testing size fixed at 6000 feature vectors (500 per class) and calculate the overall classification accuracy metric. Fig. \ref{fig:accuracy_samples} shows the classification accuracy versus training samples required per class. For the beam pattern feature, we see that with 3 APs less than $50$ training samples per class are required to achieve an accuracy of over $99\%$. For the PSD feature, the number of training samples required depends on the FFT size. For 128 FFT size, the overall accuracy saturates at $98.5\%$ at 350 training samples. For 256, 1024 and 4096 FFT size, the overall accuracy saturated at $99\%$ for 350 training samples. For training samples per class above 350, the classification accuracy did not improve for increasing training sample size. For lower training samples, increasing the FFT size significantly improves the classification accuracy. The PSD feature requires significantly more training samples per class when compared to our proposed beam pattern feature. Fig. \ref{fig:classifier_3AP_tsne} shows the 2-dimensional embedding of the output weights learned by the CNN classifier for the beam pattern feature 3 APs case. The 12 devices form 12 distinct clusters indicating the beam pattern feature is unique among devices and the classifier was able to learn the representations of the features to form distinct clusters. 

\subsection{Identification Results}
During the identification phase, the claimed identity (e.g., MAC address) of the device is authenticated by performing feature extraction and comparing it with the feature of the claimed device stored in the fingerprint database. In this section, we compare the identification performance of our proposed beam pattern feature with the PSD feature for stationary and NLOS scenarios. In addition, for the beam pattern feature we assess the impact of mobility (device orientation change) on the identification and highlight the challenges. 
\subsubsection{Stationary}
\begin{figure*}[htbp!] 
 %\centering
  \subfloat[\label{fig:roc_stationary_codebook}
]{\includegraphics[trim={0 0 2cm 1cm},clip,width=0.23\textwidth]{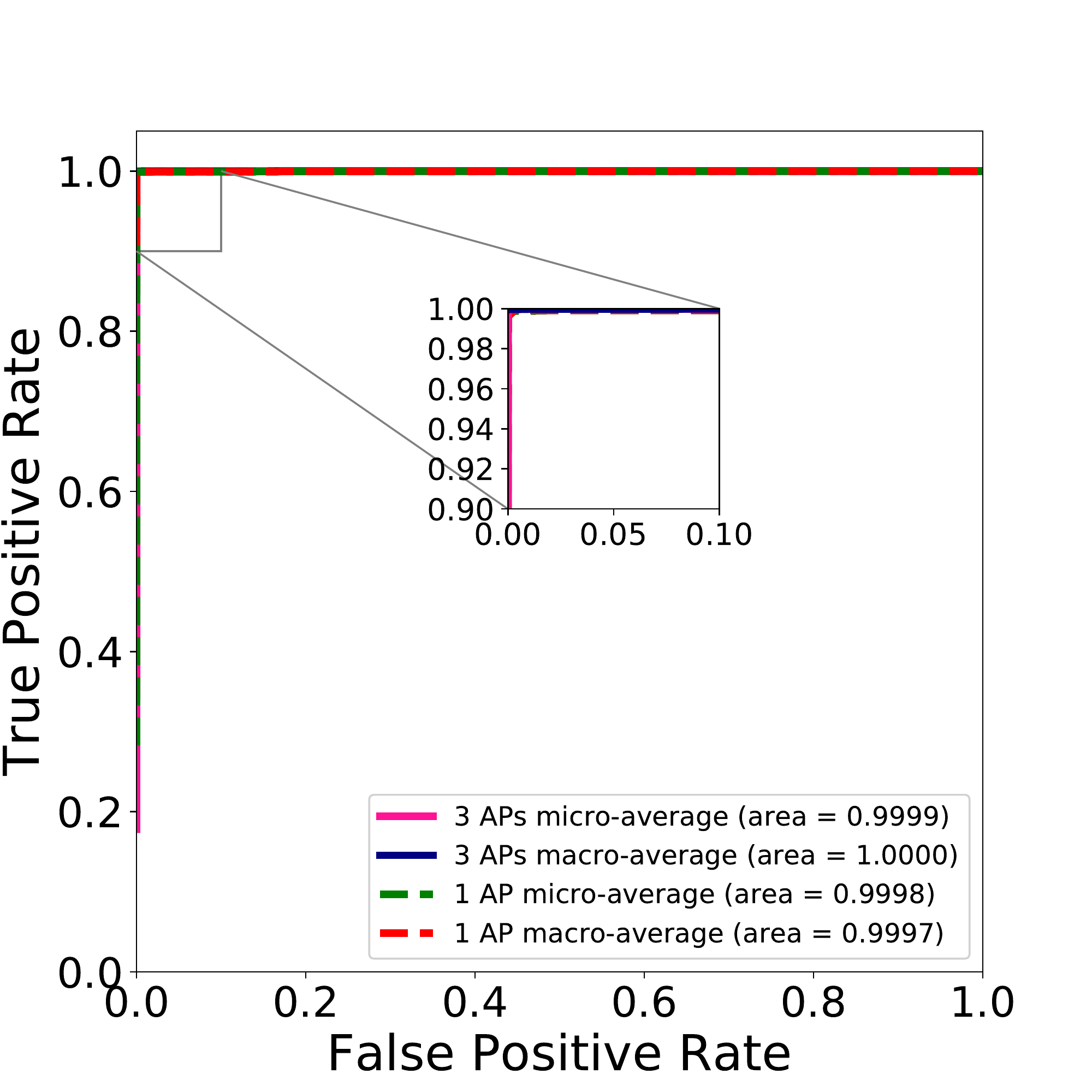}} \hspace{0.5em}
  \subfloat[\label{fig:roc_stationary_psd}
]{\includegraphics[trim={0 0 2cm 1cm},clip,width=0.23\textwidth]{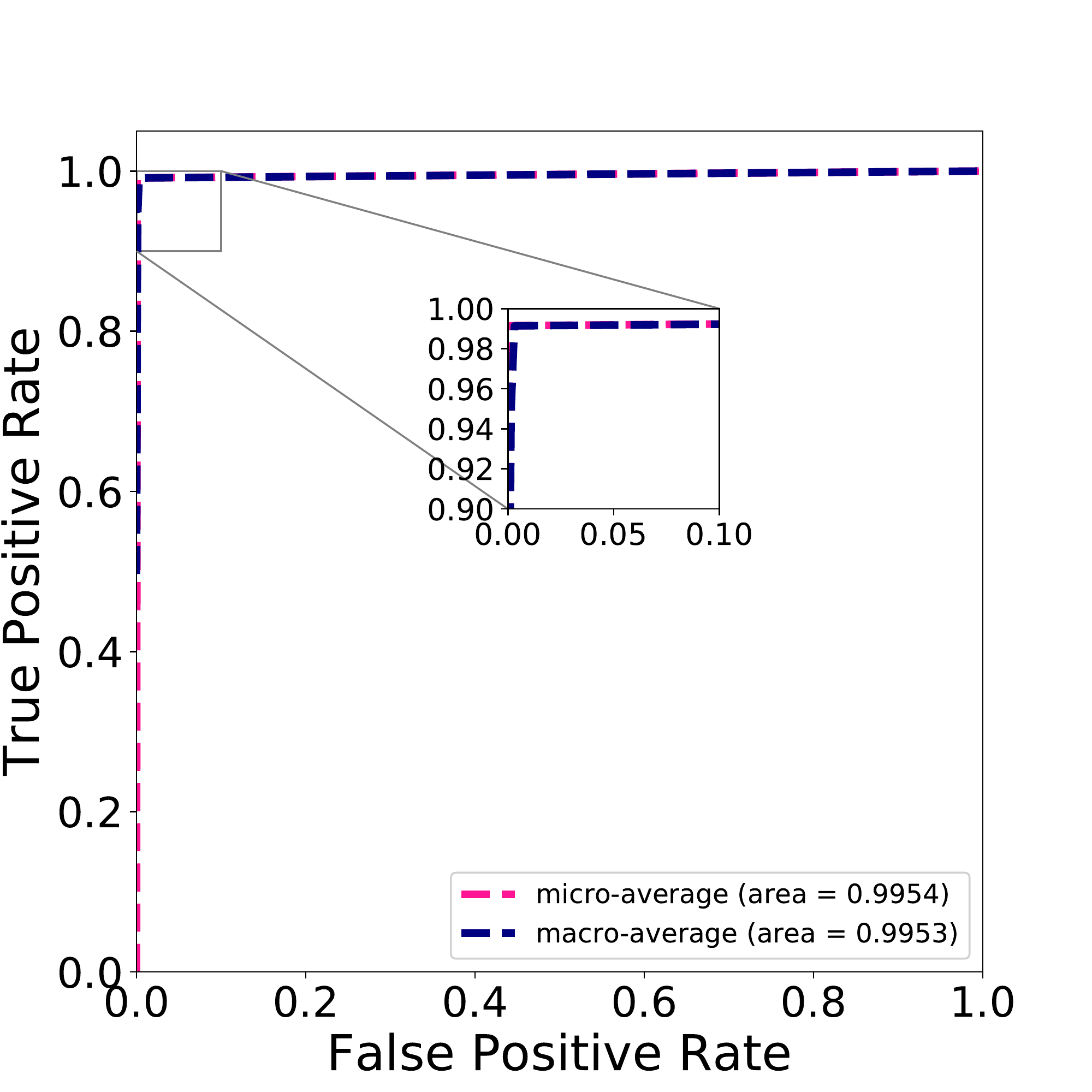}}\hspace{0.5em}
 \subfloat[\label{fig:NLOS_roc_bemapattern}]{\includegraphics[trim={0 0 2cm 2cm},clip,width=0.23\textwidth]{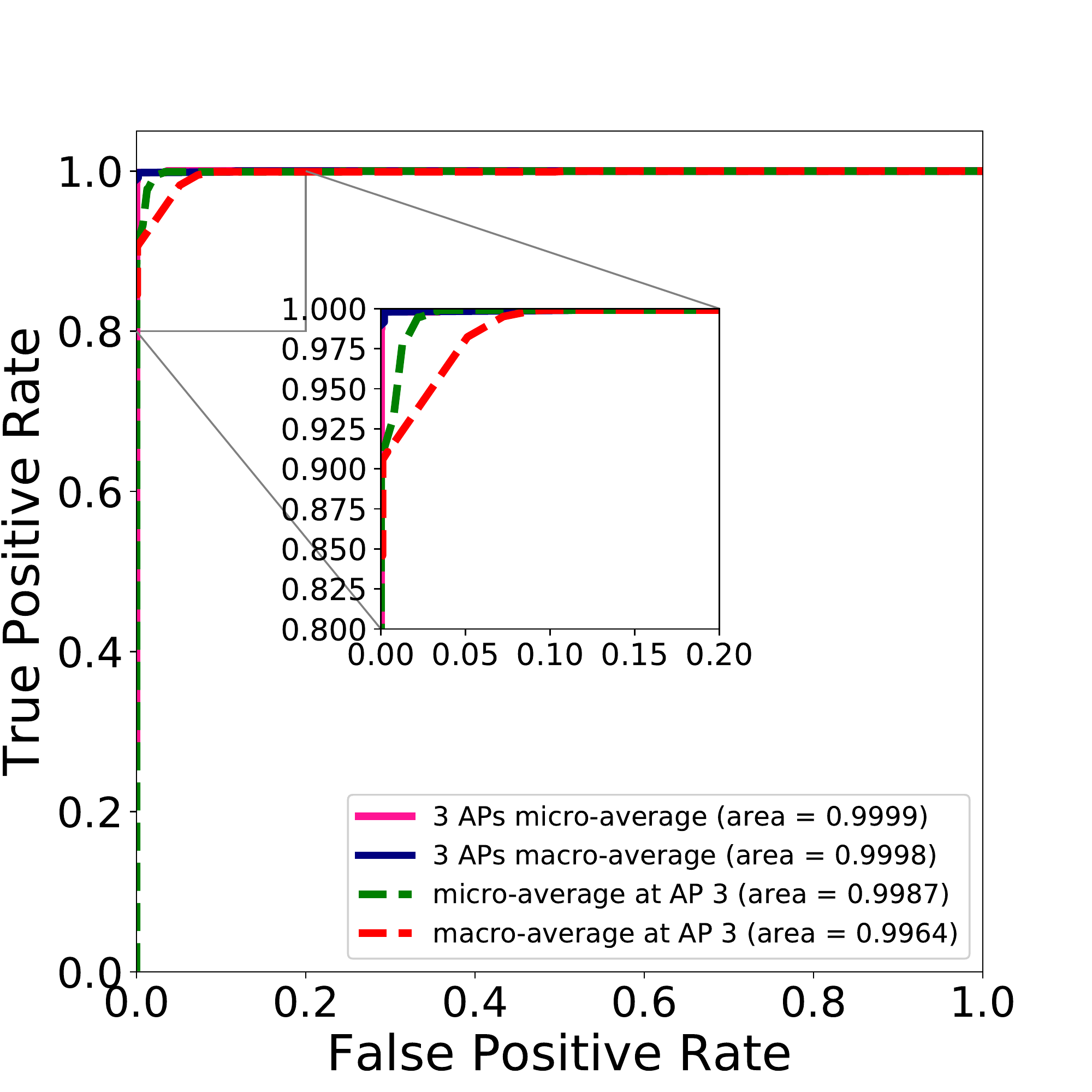}} \hspace{0.5em}
  \subfloat[\label{fig:NLOS_roc_psd}]{\includegraphics[trim={0 0 2cm 2cm},clip,width=0.23\textwidth]{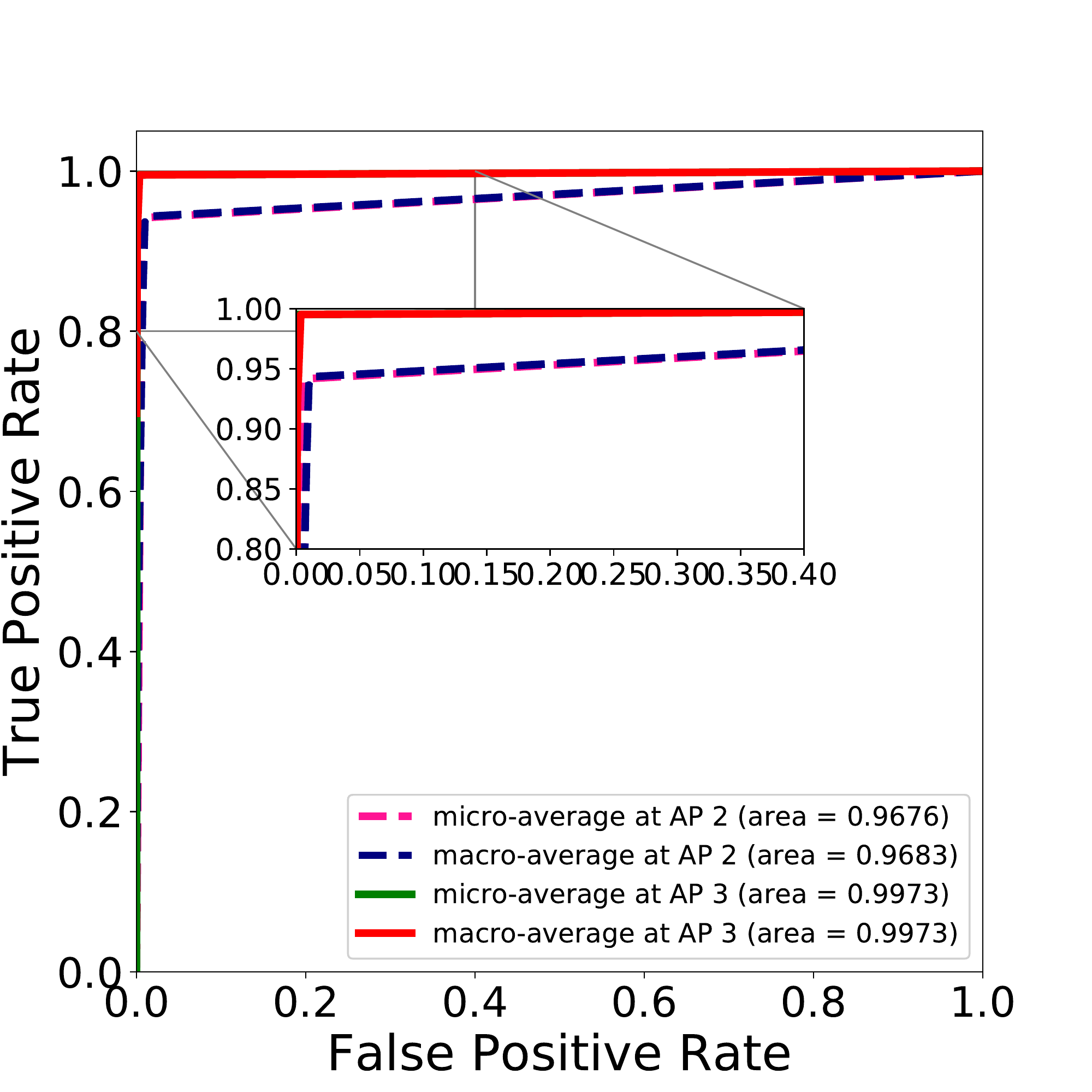}} \hspace{1em}
  \caption{ROC curves: (a) beam pattern stationary, (b) PSD stationary, (c) beam pattern NLOS and (d) PSD NLOS.}
\end{figure*}
This experiment pertains to scenarios involving communication between laptops and APs in office, conference and living room scenarios, where the devices are kept stationary with respect to their position and orientation \cite{TGayUseCases}. All the devices are tested with the same position and orientation in the location depicted as star symbol in Fig. \ref{fig:floor plan}. At each AP, a 1000 feature vector per device is obtained and split into training, validation and test sets of $35\%$, $15\%$ and $50\%$ respectively. Fig. \ref{fig:roc_stationary_codebook} shows the ROC curves for 1 AP and 3 APs scenario for the beam pattern feature. We see that the area under the ROC curve (ROC-AUC) is $0.998$ and $0.999$ for 1 AP and 3 APs case respectively. The EERs for all the devices are $<1\%$ for both 1 AP and 3 APs case. For the beam pattern feature, we see that, as long as the device is fingerprinted and identified in the same orientation, the overall accuracy is $99.6\%$ and $99.9\%$ for 1 AP and 3 APs case respectiveley. Fig. \ref{fig:roc_stationary_psd} shows the ROC curve for 4096 FFT PSD feature based identification. The ROC-AUC for 128, 256, 1024, and 4096 FFT PSD are all over $0.99$. The EERs are $<1\%$ for all the devices and all FFT size PSD features. The average identification accuracies for PSD feature with 128, 256, 1024 and 4096 FFT are $97\%$, $98.4\%$, $99\%$, and $99.1\%$ respectively. A closer look at the identification performance reveals that for 128 FFT and 256 FFT, some of the Intel based devices identification accuracy is as low as $89\%$ and $95\%$ respectively. Hence for PSD feature fingerprinting, higher FFT size on the order of 1024 and higher is needed to achieve an identification accuracy of over $99\%$, which significantly increases the complexity of the fingerprinting system when compared to beam pattern feature based system.   
\subsubsection{Effect of Channel}
In practical scenarios, the line-of-sight of the device to be identified might be blocked by obstacles in the environment and only NLOS might be available. Experiments are performed to assess the identification performance under such NLOS scenarios. The user devices are positioned at a location identified as \textit{NLOS} behind a cubicle partition as shown in Fig. \ref{fig:floor plan}. Among the 3 APs, only AP2 is in LOS with respect to the user devices. The orientation and position of all the devices are kept the same. 1000 feature vector per device at each AP is obtained. The database is split into training, validation and test set with $35\%$, $15\%$, and $50\%$ respectively. Fig.\ref{fig:NLOS_roc_bemapattern} shows the ROC curve for the beam pattern feature in NLOS scenario. Due to space constraints, we show the ROC curves only for the NLOS AP namely AP3 and 3 APs case. Only 6 devices have EER $<5\%$ for the NLOS AP 3. For the 3 APs case, the EERs of all the devices are $<1\%$. For the single AP case, we see that  the performance depends on the location of the AP. The average identification accuracy across all the devices are $95.7\%$, $99.2\%$ and $94.2\%$ at AP1, AP2 and AP3 respectively. The average identification accuracy for 3APs case is $99.5\%$. Employing multiple APs significantly increases the identification accuracy in the NLOS case. For the PSD feature, we report the metric for only 4096 FFT PSD due to space constraints. The ROC curves for PSD feature under NLOS scenario at NLOS AP3 and LOS AP2 is presented in Fig. \ref{fig:NLOS_roc_psd}. The EERs are all $<1\%$ for the LOS AP2 and $<5\%$ for the NLOS AP3. The average identification accuracies are $94\%$, $99\%$ and $94\%$ at AP1, AP2 and AP3 respectively. We see that, the beam pattern feature is robust to NLOS and the identification accuracy is significantly higher than the PSD feature in the NLOS scenario.

\subsubsection{Effect of Mobility}
In this section, we seek to answer the following question: \textit{How much does the mobility of the device affect the performance of the beam pattern based fingerprinting system?} We categorize the experiments into limited mobility and high mobility according to the user mobility behavior. For each category, we conduct comprehensive experiments to understand their impact on the identification system. The mobility model is adopted from IEEE 802.11ay channel model\cite{maltsev2015channel}. The mobility of the device can be represented by velocity vector $v=v_{x}\dot i+v_{y}\dot j+v_{z}\dot k$ which is the sum of the scalar components along $x, y, z$ directions. For our mobility experiments, the vertical component of the velocity $v_z$ is 0 meaning the devices are kept at the same height. Due to mobility, the user device performs beamforming training to find or update its best direction (sector) to communicate. The beam pattern feature vector is extracted from the beam searching beacons received along the mobility path. For each device, 500 beam pattern feature vector per device is obtained along the mobility path. 10 devices (Talon1 to 3, Acer1, Netgear1 to 2, and Intel1 to 4) are evaluated in total.   
\begin{figure}[htbp!]
\subfloat[  \label{fig:straight_walking_roc}
]{\includegraphics[trim={0 0 2cm 1cm},clip,width=0.23\textwidth]{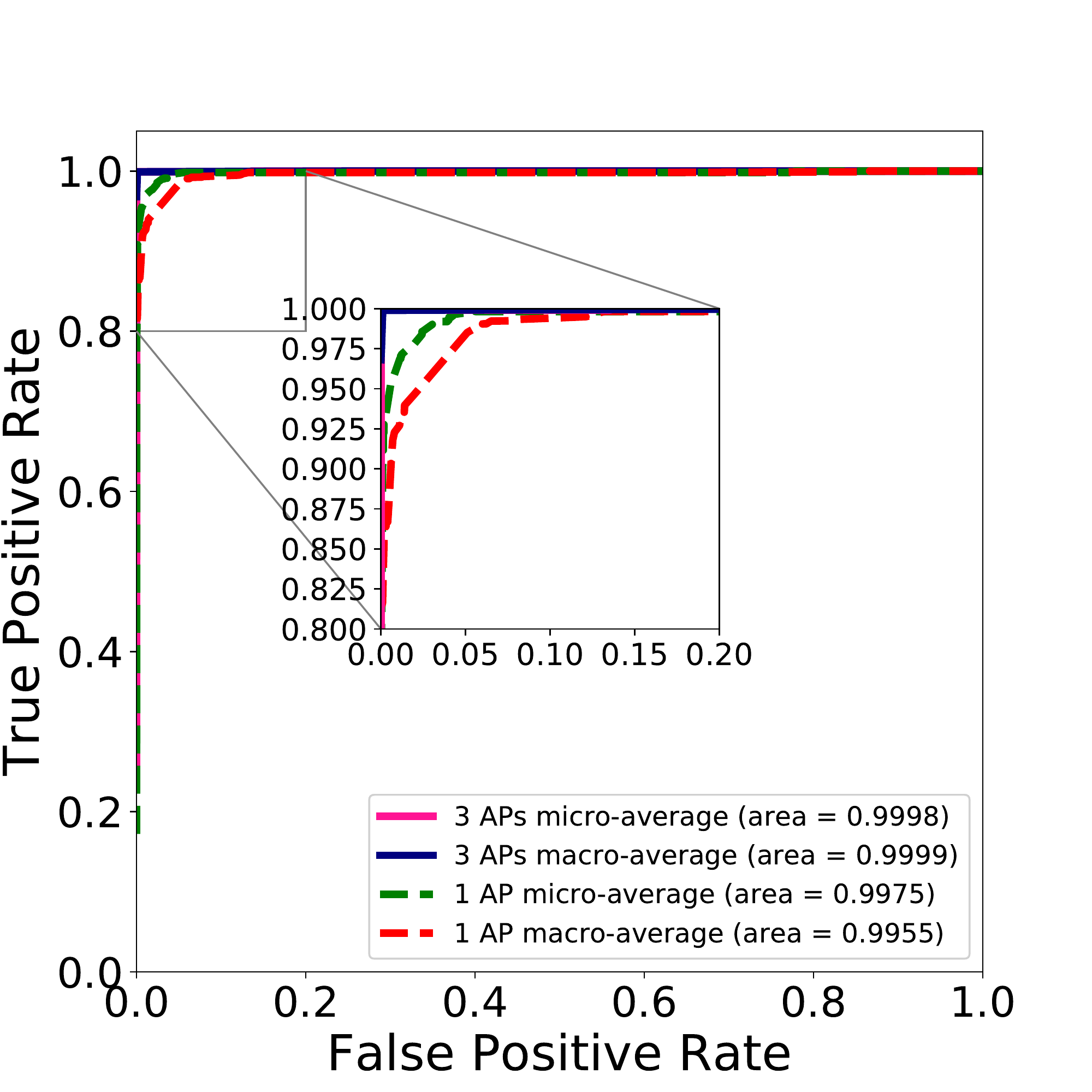}}\hspace{0.5em}
\subfloat[\label{fig:high_mobility_1AP_roc}
]{\includegraphics[trim={0 0 2cm 1cm},clip,width=0.23\textwidth]{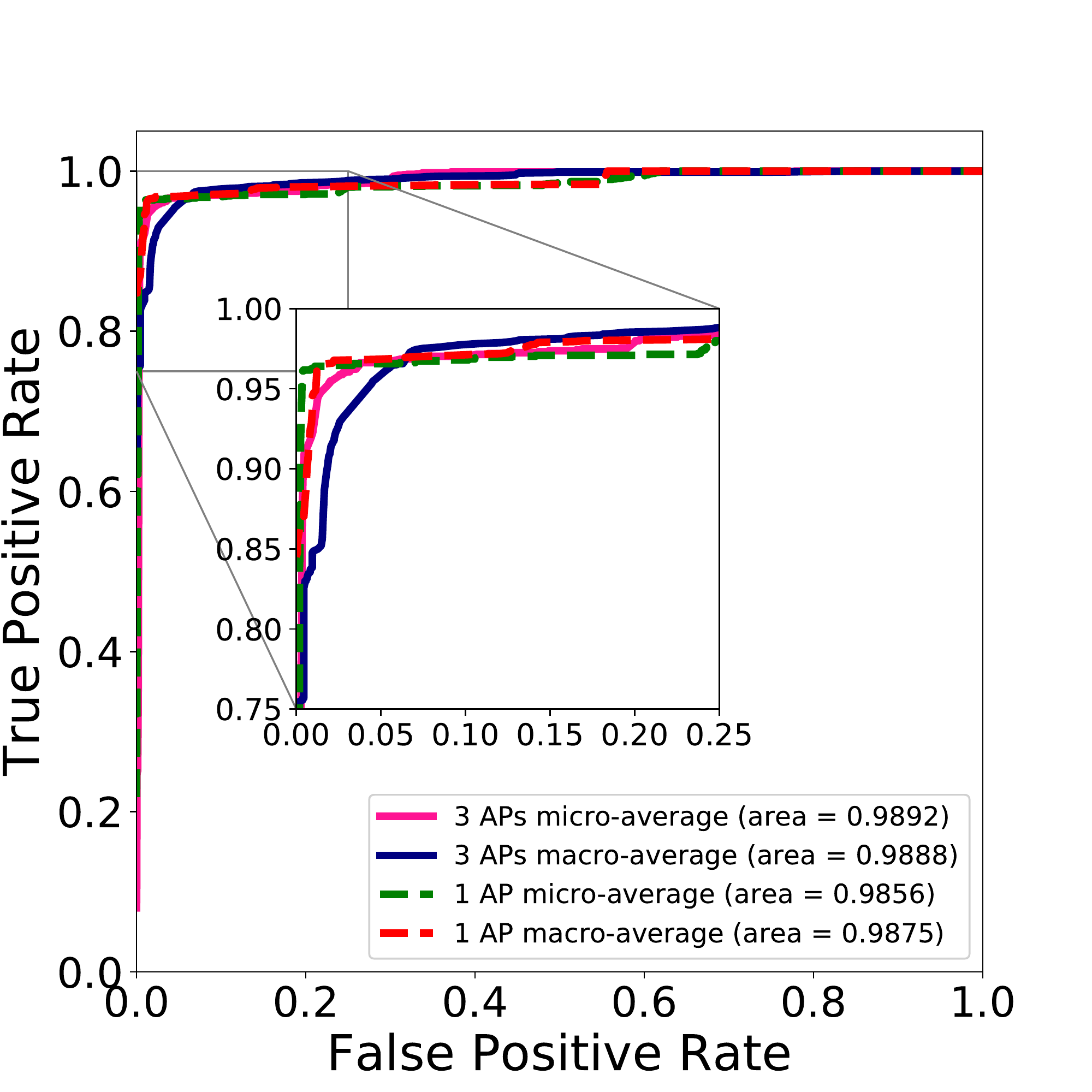}}
\caption{ROC curves: (a) limited mobility and (b) high mobility.}
\end{figure}
\paragraph*{\textbf{Limited Mobility}}
We specifically look at the scenario where the user is walking straight with limited device orientation change. Assuming the user is moving along the $x$ direction, the movement along the $y$ direction (velocity vector $v_y$) is kept minimum to imitate a user walking straight with holding the device steady. The devices are kept in a cart and moved with walking speed (1-2m/s) along the path shown in the Fig. \ref{fig:floor plan}. Throughout the experiment, the device orientation is kept constant. Fig. \ref{fig:straight_walking_roc} shows the ROC curves. The EERs of all the devices are $<10\%$ and $<1\%$ for 1 AP and 3 APs case respectively. The average accuracies for 1 AP (at AP1) and 3 APs case are $95\%$ and $99.7\%$ respectively. This shows that obtaining multiple views of the beam pattern through multiple APs significantly increases the identification accuracy.  
\paragraph*{\textbf{High Mobility}}
Here we look at the mobility scenario where the direction of mobility and orientation are random. The device is moved to imitate walking speed of human. This experiment captures a real-world practical scenario and tests the limitations of our fingerprint database building method as random device orientation change introduced by mobility may result in the beam pattern feature vector not being represented in the feature training database. Fig. \ref{fig:high_mobility_1AP_roc} shows the ROC curve for this scenario for 1 AP and 3 APs fingerprinting system. For a 1 AP system, only six (Talon1, Talon3, Acer1, Netgear1, Netgeat2, Intel3) of the 10 devices achieve an EER$\leq 5\%$. On the other hand, for 3 APs system, all but Intel4 achieve an EER$\leq 5\%$. The overall identification accuracy for 1 AP and 3 AP system is $92\%$ and $96\%$ respectively showing significant advantage of utilizing multiple APs system for mmWave device fingerprinting.     

\paragraph*{\textbf{Impact of Mobility on Identification}}\label{sec:effect of mobility}
\begin{figure}[!htbp]
 \centering
 \includegraphics[trim={0 0 0cm 0 },clip,width=0.25\textwidth]{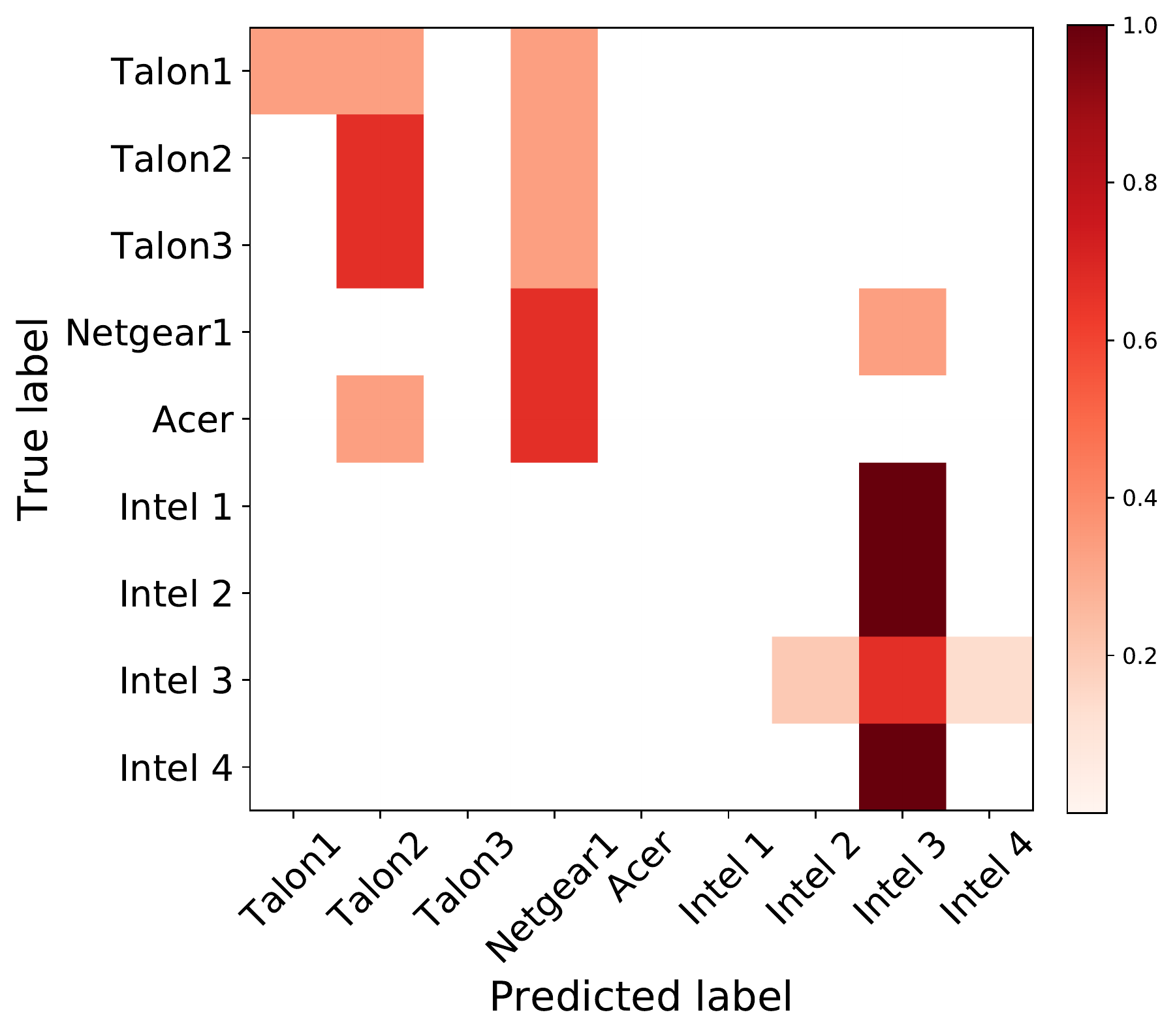}
 \caption{Confusion matrix for beam pattern feature for orientation change.}
 \label{fig:mobility_worstcase}
\end{figure}
The identification accuracy of the beam pattern feature based identification system depends on the orientation of the device to be identified. If the device orientation changes, different from the training fingerprint, the accuracy drastically decreases. We looked at the impact of device mobility (orientation change) on a beam pattern feature fingerprinting system by keeping only the fingerprints extracted from one fixed direction in the database and excluding all others. The devices are identified from a direction other than the one in the trained database. Fig. \ref{fig:mobility_worstcase} shows the confusion matrix for such a scenario. The fingerprinting system could not discriminate the devices and the average accuracy is $25\%$. Hence, it is essential to learn as many orientations in the beam pattern as possible through device mobility (see Sec. \ref{sec:learning procedure}). 
\section{Security Analysis And Experimental Evaluation}
\label{sec:attacks}
It is generally believed that physical layer security which relies on unique device dependent features generated by hardware imperfections of the RF chain and the antenna module is hard to forge. However due to the availability of high end software defined radios and waveform generators,  physical layer features are prone to attacks such as feature and signal replay. Also, the novel beam pattern feature proposed in section \ref{sec:feature origin} introduces an additional attack vector unique to mmWave wireless systems. 
\par
In this section, we study the security of our proposed beam pattern based device identification method and PSD feature based device identification against impersonation attacks. 
\paragraph*{\textbf{Threat Model}}
\label{sec:Adversarial capabilities}
\begin{figure}[htbp!]
\subfloat[\label{fig:NI beam pattern}]{\includegraphics[trim={0 0cm 0cm 2cm},clip,width=0.24\textwidth]{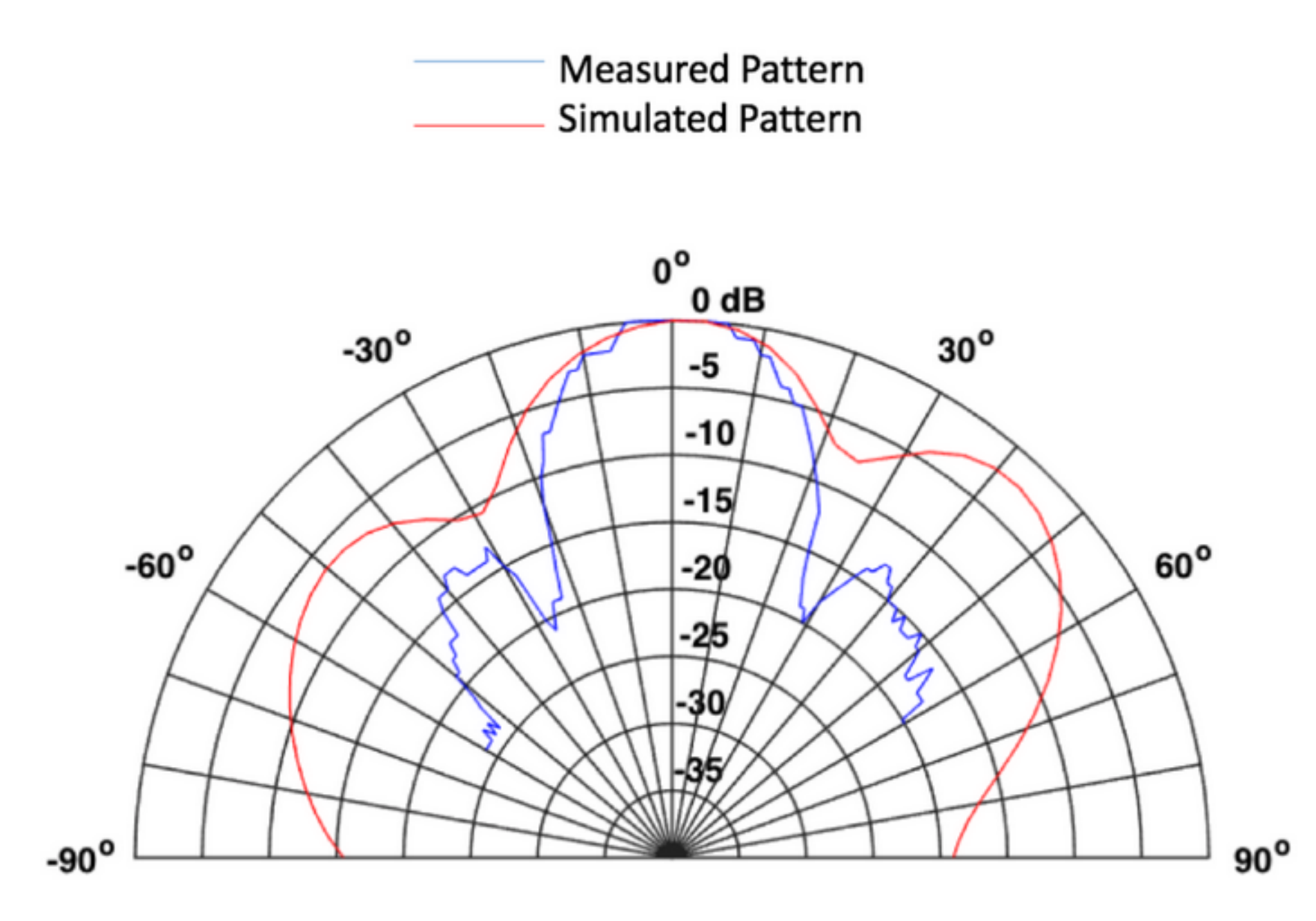}} %\hspace{2em}
\subfloat[]{\includegraphics[trim={1cm 6cm 2cm 6.5cm},clip,width=0.24\textwidth]{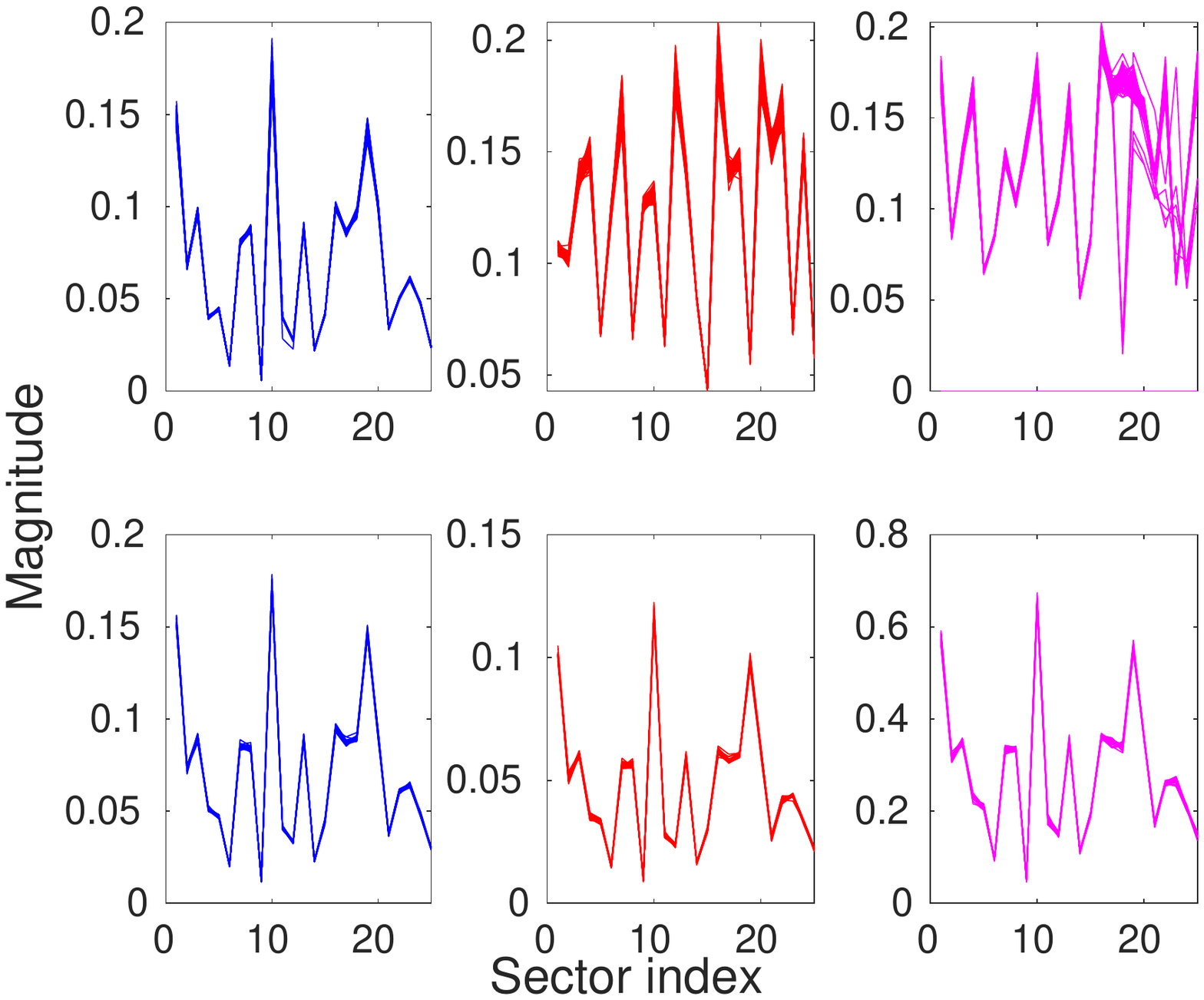}}
\caption{(a) X60 beam pattern 12 and (b) beam pattern feature of \textit{Bob} at 3 APs (top row) and impersonated beam pattern feature at 3 APs (bottom row).}
\end{figure}
The goal of \textit{Mallory} $\mathcal{M}$ is to gain access to the wireless network by impersonating the identity of a legitimate user \textit{Bob} in the network. \textit{Mallory} may achieve this goal in 2 stages: (1) \textit{eavesdropping}, and (2) \textit{signal impersonation}.
\par Before proceeding to describe the attack stages, we make the following assumptions about \textit{Mallory}: (a) \textit{Mallory} possesses a millimeter wave RF transceiver (Sec \ref{sec:Attacker device}) and has the knowledge of the communication channel (60.48 GHz in our experiments) used by \textit{Alice} and \textit{Bob}. The active communication channel could also be deduced by RF sensing the environment. (b) \textit{Mallory} can use both directional antenna and quasi-omni antenna for eavesdropping and impersonation attacks. (c) We assume \textit{Mallory} does not know the classification/identification method used by the wireless network. 
\par During the eavesdropping stage, \textit{Mallory} records the beam searching beacons of \textit{Bob} whose identity she aims to fake using quasi-omni antenna.
\iffalse
 \textit{Mallory} has the choice of using either directional or quasi-omni antenna for eavesdropping. If \textit{Mallory} knows the location of \textit{Bob}, she can use the directional antenna to steer her receive beam towards \textit{Bob}. Using directional antenna will enable \textit{Mallory} to record \textit{Bob's} beacon with high signal-to-noise ratio or she can record the beacons from further distance than using quasi-omni antenna.
 However, using directional antenna will make it difficult for \textit{Mallory} to receive beacons that are not directed towards her.  Therefore, in our scenario, \textit{Mallory} uses quasi-omni antenna. 
 \fi
  \textit{Mallory} could simply record the beam searching beacons of \textit{Bob} as long as she is in the communication range of \textit{Bob}. 
\par Followed by the eavesdropping stage, \textit{Mallory} initiates signal impersonation stage. \textit{Mallory} replays the recorded beacons from \textit{Bob} either unchanged or after modifying the signal. Here too \textit{Mallory} can choose between a directional and quasi-omni antenna pattern. During the impersonation stage, \textit{Mallory's} objective is to replay the signal towards the access point \textit{Alice}. Here we make the assumption that \textit{Mallory} knows the location of \textit{Alice} and hence can direct her transmit beam towards \textit{Alice} using directional beam pattern. 

\subsection{Impersonation of Antenna Pattern Feature}
\label{sec:AntennaPatternAttack}
In this section, we describe the antenna pattern feature impersonation attack.  
\paragraph{Experimental Setting}
\label{sec:Attack model}

We used the X60 node described in Sec. \ref{sec:Attacker device} to capture the beacons from the legitimate device \textit{Bob}. The receiver's antenna is configured in the quasi-omni mode for capturing the beacons. The fingerprinting AP positions are as shown in Fig. \ref{fig:floor plan}. The legitimate device is placed at a position 3 meters from the AP 1. The attacker's receiver is placed directly behind the fingerprinting AP 1 to capture the beacons. The captured baseband complex samples are then transfered to the host computer and saved in a file. 
The captured beacons are read from the file and transfered from the host computer to the FPGA in the X60 used for replay. The baseband I/Q samples are then upconverted to IF and then to 60.48 GHz. The attackers transmitter is configured to use directional antenna with beam pattern shown in Fig.\ref{fig:NI beam pattern}. The beacons are replayed from a position behind the legitimate transmitter \textit{Bob}.  

\paragraph{Attack on Classification and Identification System}

\begin{figure}[htb!]
\subfloat[\label{fig:1 AP beam pattern attack}]{\includegraphics[trim={0cm 0cm 0cm 0cm},clip,width=0.24\textwidth]{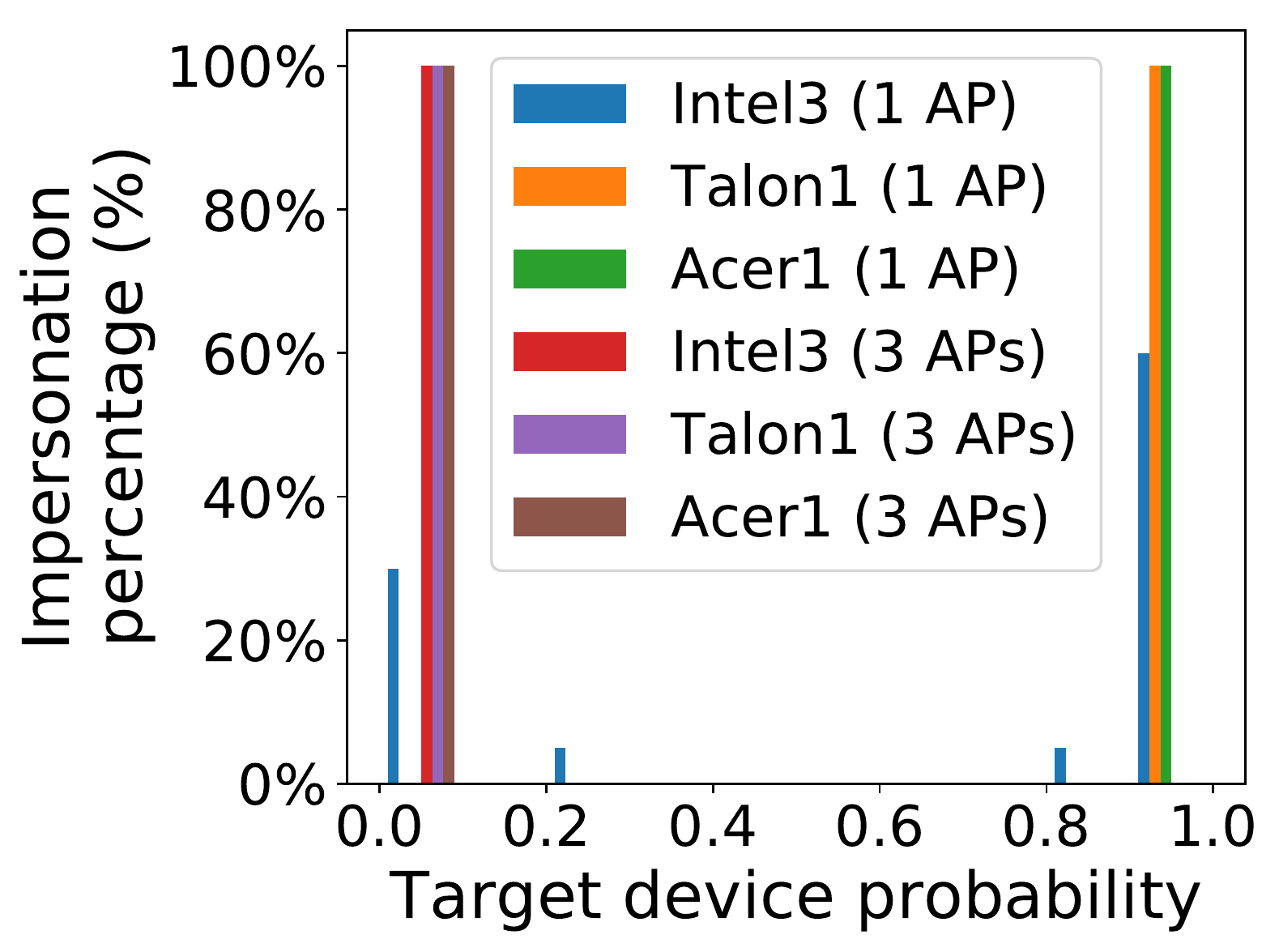}}
\subfloat[\label{fig:psd_attack}
]{\includegraphics[trim={0 0cm 0cm 0cm},clip,width=0.24\textwidth]{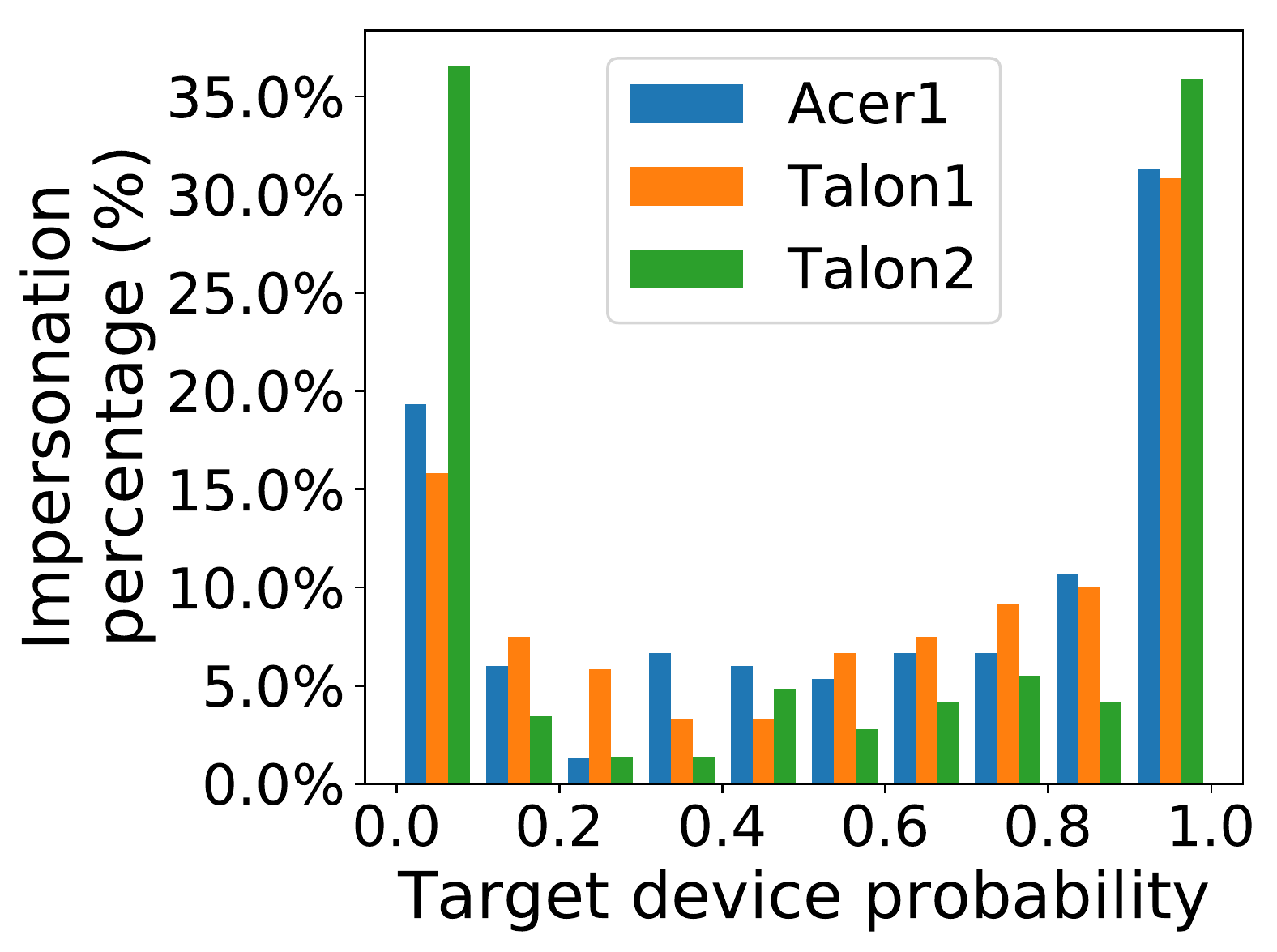}}
\caption{Signal replay attack: (a) beam pattern feature and (b) PSD feature.}
\end{figure}
Here we look at the security implications of antenna pattern feature signal replay attack on our proposed classification and identification system. In particular we explore whether the attacker was able to spoof the classifier in assigning the beam pattern feature extracted from the replayed signals to the targeted device class. For this experiment, the classifier in Sec. \ref{sec:cnn} is trained with training data from all the 12 devices. We choose three of the 12 devices (Talon1, Acer1, and Intel3) as the target devices and replay 250 sets of beam searching beacons from each of the target device using the attacker. We record the replayed beacon frames from the attacker at the three AP positions. The replayed frames are passed to the trained classifier and the identification accuracy is evaluated. The attack is evaluated for both 1 AP and 3 APs case. Fig. \ref{fig:1 AP beam pattern attack} show the percentage of frames and their associated target probabilites assigned by the classifier for three target devices for both 1 AP and 3 APs case. For the Talon1 and Acer1 attacks, we find that for single AP scenario, the attacker was successfully able to spoof the classifier with $100 \%$ accuracy and more importantly the replayed frames are classified with $99\%$ probability as those belonging to the targeted device. But for Intel3 device attack, $60\%$ and $5\%$ of the replayed frames are assigned with probabilites 0.9 and 0.8 respectively as targeted device Intel3. The remaining $35\%$ of the replayed frames are misidentifed as some other device. The overall attack success rate for Intel3 device is $65\%$. We found that the beam pattern vector extracted from the replayed frames deviates from the beam pattern of the legitimate device at few positions in the extracted feature vector due to the attacker transmitter not being able to accurately represent those beacons. To improve the attack success rate in such scenarios, we adopted an approach similar to the hill-climbing attack \cite{uludag2004attacks} in biometric systems. In a hill-climbing attack, the attacker repeatedly replays the signal each time with slight modifications until the system accepts it as genuine. Similar approach could be used to attack our beam pattern feature by crafting the beacons to be replayed. Implementing such an attack for our beam pattern replay is a daunting task as the feature vector is high dimensional with length depending on the number of codebooks used by the device. Therefore, we designed a coordinated attack in which we placed another attacker receiver behind the AP we want to impersonate. The attacker receiver has a copy of the signal to be replayed. The attacker receiver records the replayed signal from the attacker transmitter and compares with the one transmitted. In that way the attacker knows which beacon signal need to be crafted to improve the attack success rate. Using the signal crafting approach, the success rate for attack on Intel3 is $100\%$. As mentioned in \cite{danev2010attacks}, the success of the attacker to impersonate a target device in identification accuracy highly depends on the number of devices used. For larger number of devices, the classification boundary becomes obscure and the percentage of attacker replayed signals successfully assigned to the target device will decrease. 
\par In practical scenarios, the attacker could only target 1 AP to spoof at a time. Figure \ref{fig:1 AP beam pattern attack} show the percentage of frames and their associated target probabilites assigned by the classifier for three target devices for the 3 APs case. We see that for all the three targeted devices, the replayed frames are assigned probabilites $<1\%$ as targeted device and the attacker was not able to spoof the 3 APs fingerprinting system. 
\subsection{Impersonation of Spectral Feature}
\label{sec:PSDAttack}
In this section, we discuss the impersonation of the spectral feature for mmWave devices discussed in Section \ref{sec:psd}. 
\paragraph{Experiment Setting}
In our spectral feature impersonation experiment, the attacker records the beacons and replays them in a set up same as in Sec. \ref{sec:AntennaPatternAttack}. 
\paragraph{Attack on Classification and Identification System}
The identification system is trained with the classifier in Sec. \ref{sec:cnn} and the training set consists of a spectral feature fingerprint database from all the 12 devices. Four different spectral feature identification systems are considered: spectral feature with 128 FFT, 256 FFT, 1024 FFT, and 4096 FFT. The spectral feature impersonation attack is performed for each of the identification system. The impersonation attack is targeted for three legitimate devices named Acer1, Talon1, Talon2 from the enrolled devices. For each device, a total of 250 frames are replayed. The spectral feature for different FFT length is extracted from the 250 frames and submitted to the identification system. Figure \ref{fig:psd_attack} show the output probabilities assigned to the replayed frames for 4096 FFT spectral feature impersonation attack on Talon1, Talon2 and Acer1 devices. Over $30\%$ of the replayed frames are accepted as genuine frames with high probability over 0.9. The overall impersonation success rate for 4096 FFT spectral feature attack are $63\%$, $64\%$ and $53\%$ for Acer1, Talon1 and Talon2 respectively. The impersonation success rate for impersonation attack on 128 FFT, 256 FFT and 1024 FFT spectral feature are $42\%$, $43.3\%$, and $50\%$ respectively for Acer1 device. For Talon1, they are $35\%$, $35.8\%$, and $53.3\%$ for 128 FFT, 256 FFT, and 1024 FFT respectively. For Talon2, they are $30\%$, $31\%$ and $41\%$ for 128 FFT, 256 FFT and 1024 FFT respectively. We noticed that as the FFT resolution goes down, the success rate of attack also lowers. This trend is observed for all the devices we impersonated. The reason for this is that, with lower FFT resolution feature, the classification and identification accuracy also reduces. This leads to reduction in attack success rate as some of the replayed frames are incorrectly identified as devices other than the one impersonated.    
\subsection{Security Implications}
Our findings suggest that similar to the PSD feature and other conventional features \cite{danev2010attacks}, our proposed beam pattern feature too is vulnerable to impersonation attacks. However, by utilizing multiple views of the beam pattern to authenticate a legitimate user, such attacks can be defeated. We see that the beam pattern feature with multiple views has higher resilience to impersonation attacks than the PSD feature. Robustness of beam pattern feature to impersonation attacks has several implications in mmWave wireless networks security. MmWave devices could be authenticated using beam pattern physical layer feature during the beam searching process. Thus, attacks such as \cite{Steinmetzer2018BeamStealingIT} could be prevented by accepting the sector sweep feedback frame from the device only if it has been authenticated during the beam searching phase.

\section{Conclusion}\label{sec:conclusions}
In this paper we presented a novel beam pattern feature for fingerprinting mmWave devices. The proposed beam pattern feature is contributed by the fabrication process of the antenna array and the phase shifters used in the mmWave devices. We showed that, the directionality of mmWave devices poses additional significant challenges in the training as well as in the identification process. Comprehensive training is necessary to achieve an acceptable classification and identification accuracy under mobility conditions. To that end, we proposed a multiple APs fingerprinting architecture that exploits the rich spatial features of the beam patterns used by the mmWave devices. We also presented a conventional PSD feature for mmWave devices and compared our proposed beam pattern feature with it. We performed extensive experiments under various real-world scenarios to confirm the effectiveness and reliability of the proposed beam pattern feature. 
\par We have also investigated the robustness of our proposed beam pattern feature to impersonation attacks. We designed and implemented our attacker on a 60 GHz mmWave testbed and performed signal replay attack on the proposed beam pattern and spectral features. We show that the impersonation attack on the beam pattern feature for a single AP system is almost always feasible with high success rate while for the multiple APs system it is unsuccessful. For the spectral feature impersonation attack, the degree of success varies highly from $40\%$ to $60\%$. This suggests that the beam pattern feature is robust and secure when compared to a conventional PSD based scheme. 
% or


\appendices



\ifCLASSOPTIONcaptionsoff
  \newpage
\fi



\bibliographystyle{IEEEtran}
\bibliography{refs}


%\vfill


\end{document}